\documentclass[sigconf]{acmart}

\usepackage{multirow}
\usepackage{colortbl}
\usepackage{pifont}
\newcommand{\cmark}{\ding{51}}
\newcommand{\xmark}{\ding{55}}
\usepackage{enumitem}
\usepackage{cleveref}
\usepackage{balance}

\AtBeginDocument{%
  }

\copyrightyear{2025}
\acmYear{2025}
\setcopyright{cc}
\makeatletter
\def\@ACM@copyright@check@cc{} 
\makeatother
\setcctype{by}
\acmConference[CHI '25]{CHI Conference on Human Factors in Computing Systems}{April 26-May 1, 2025}{Yokohama, Japan}
\acmBooktitle{CHI Conference on Human Factors in Computing Systems (CHI '25), April 26-May 1, 2025, Yokohama, Japan}\acmDOI{10.1145/3706598.3713963}
\acmISBN{979-8-4007-1394-1/25/04}

\begin{document}

\title{Access Denied: Meaningful Data Access for Quantitative Algorithm Audits}

\author{Juliette Zaccour}
\email{juliette.zaccour@oii.ox.ac.uk}
\affiliation{%
  \institution{Oxford Internet Institute,\\University of Oxford}
  \city{Oxford}
  \country{United Kingdom}
}

\author{Reuben Binns}
\email{reuben.binns@cs.ox.ac.uk}
\affiliation{%
  \institution{Department of Computer Science, University of Oxford}
  \city{Oxford}
  \country{United Kingdom}
}

\author{Luc Rocher}
\email{luc.rocher@oii.ox.ac.uk}
\affiliation{%
  \institution{Oxford Internet Institute,\\University of Oxford}
  \city{Oxford}
  \country{United Kingdom}
}

\begin{abstract} 
Independent algorithm audits hold the promise of bringing accountability to automated decision-making. However, third-party audits are often hindered by access restrictions, forcing auditors to rely on limited, low-quality data. To study how these limitations impact research integrity, we conduct audit simulations on two realistic case studies for recidivism and healthcare coverage prediction. We examine the accuracy of estimating group parity metrics across three levels of access: (a) aggregated statistics, (b) individual-level data with model outputs, and (c) individual-level data without model outputs. Despite selecting one of the simplest tasks for algorithmic auditing, we find that data minimization and anonymization practices can strongly increase error rates on individual-level data, leading to unreliable assessments. We discuss implications for independent auditors, as well as potential avenues for HCI researchers and regulators to improve data access and enable both reliable and holistic evaluations.
\end{abstract}



\received{12 September 2024}
\received[revised]{10 December 2024}
\received[accepted]{16 January 2025}

\maketitle

\section{Introduction}

Automated decision-making and decision-supporting systems are becoming commonplace across society, with algorithms used in criminal justice, healthcare, social welfare, child protection, and immigration sectors. Despite their widespread use, these systems are commonly deployed without external and independent oversight. In the past few years, researchers and journalists have investigated high-stakes yet highly secretive algorithmic decisions such as visa applications and welfare benefits applications~\cite{maxwell2022,vanbekkum2021}. Such algorithms have been shown to cause significant harm and disproportionately affect marginalized individuals, at a larger scale, with lower accountability than human decision-making~\cite{barocas2023}.

Algorithm audits, defined as evaluations of algorithmic systems for accountability purposes~\cite{birhane2024}, have been an important avenue through which researchers have shed light on algorithmic harms, often sparking critical conversations around algorithmic justice both within the field and in media~\cite{angwin2016,obermeyer2019,chouldechova2016,lum2016}. Third-party audits---as opposed to first- and second-party audits performed by internal or contracted auditors---are widely considered an efficient and necessary form of oversight and source of accountability across industries~\cite{costanza-chock2022a,raji2022}. While there are examples of third-party algorithm audits~\cite{seidelin2022,obermeyer2019}, successes are relatively few. This is largely due to significant challenges preventing adequate access~\cite{holstein2019,ojewale2024,mokander2021}.\\

Effective oversight is often hindered by limited data access~\cite{trask2024}. High secrecy prevents civil society, journalists, and researchers alike from gathering basic information about algorithmic systems for qualitative evaluations, let alone individual-level data for quantitative audits. Several regulating organizations such as the European Union Commission~\cite{europeancommission2021,europeanunion2024} and the White House~\cite{thewhitehouse2023} have called for algorithm audits. In the European Union (EU), the Digital Markets Act and Digital Services Act have started to implement audit requirements for algorithmic systems. However, such provisions for audits typically target online platforms over other decision-supporting algorithms such as in public services, and tend to be unspecific and agnostic to the form of access~\cite{casper2024}. While long-established digital regulators (e.g. data protection authorities) may in theory have the legal powers to conduct such audits, they may lack the resources and typically opt for other forms of investigation in practice~\cite{adalovelaceinstitute2021a}. In the absence of legal precedents, regulatory standards, or incentives for transparency, data access is granted primarily on the terms of the auditee~\cite{raji2022} rather than based on research needs. In turn, this limits the legitimacy, efficiency, and impact of audits. Robust audits require for this tendency to be reversed, whereby the access is based on audit requirements.

However, what constitutes an appropriate audit dataset is still ill-defined. Issues faced by auditors include low sample sizes~\cite{ji2020}, missing predictors~\cite{tan2018}, or the absence of individual-level data altogether~\cite{kayser-bril2021,laquadraturedunet2023}. In addition, third-party auditors face increasing challenges for privacy-preserving access to sensitive data. A range of methods have been proposed as general solutions for data sharing, including traditional data minimization and aggregation, as well as modern privacy-enhancing technologies such as differential privacy and synthetic data~\cite{cummings2024a,galdonclavell2020,gadotti2024a,beduschi2024,edelson2023}. However, preserving privacy while maintaining data utility is challenging, and the impact of these techniques on audit integrity has yet to be evaluated. Without appropriate benchmarks and standards, auditors will likely continue to receive low-quality data that are not appropriate for their purposes. We contend that solving these challenges to data access is a necessary step towards formalizing policy for algorithm auditing.

Lack of access is one example of the real-world limitations on algorithmic auditing that are often left un-addressed within purely technical algorithmic fairness research. However, such limitations are studied within more socio-technical and human-centric research which both draws from and heavily overlaps with prior and ongoing work in human-computer interaction. In addition to specialist research communities like ACM Fairness, Accountability and Transparency, such work has been published in HCI venues such as CHI (e.g.~\cite{binns2018,lee2021a,holstein2019}) and CSCW (e.g.~\cite{shen2021,hussein2020}). Work at this intersection reflects the variety of methods and approaches in HCI, and can include the more traditional user study-based work --- including studies of user perceptions of algorithmic fairness~\cite{binns2018} and folk theories~\cite{shen2021} --- as well as assessments of how tools and techniques proposed within algorithmic fairness literature might actually work for practitioners (e.g.~\cite{holstein2019,lee2021a,imana2023}). In this work, we seek to contribute to the latter strand of research, by assessing the feasibility of algorithmic fairness auditing approaches given realistic data access constraints faced by practitioners.\\

Here, we examine how to accurately audit algorithmic systems while addressing legitimate privacy concerns for individuals. We set out to identify which data sharing practices are appropriate for audits and which practices can mislead auditors. We conduct simulations of audits on two real-world datasets, across three levels of access: (a) aggregate statistics only, (b) individual-level data with model outputs, and (c) individual-level data without model access. We focus on a simple and realistic audit case study, where an auditor estimates group parity metrics for a binary classification model. We apply this audit task to two models, one predicting recidivism and one predicting healthcare coverage. We examine how loss in audit data quality affects the reliability of quantitative fairness assessments for Machine Learning (ML) classification models.

When auditing decision-making models, group parity metrics are often used as a diagnostic tool and starting point by researchers~\cite{angwin2016,obermeyer2019}. These metrics allow to estimate the real-world impact of model decisions, and their reliability is therefore critical~\cite{kallus2022,ojewale2024}. Our findings suggest that despite considering this simple diagnostic task, current data access mechanisms may compromise the integrity of auditor assessments. We show that the data quality standards necessary to ensure audit reliability are already difficult to obtain, highlighting the urgency to increase auditor access. We discuss avenues for HCI researchers and regulators to work towards effective data access for auditors.

\section{Background}

\subsection{Scoping the AI audit landscape}
Several works have recently assessed the AI audit landscape, and identified data access as an important delineation between types of algorithm audits. Birhane et al.~\cite{birhane2024} define a typology of audits, distinguishing four categories: \textit{model} or \textit{algorithm audits} where auditors diagnose biases and errors (e.g.~\cite{angwin2016}); \textit{data audits} where auditors target a dataset (e.g.~\cite{buolamwini2018}); \textit{ecosystem audits} where auditors examine socio-technical environments (e.g.~\cite{brown2021}); and \textit{meta-commentaries} for discussions about the practice of auditing and methods (e.g.~\cite{goodman2022}). While more comprehensive, ecosystem audits require a broad access to the system and organization, which is rarely possible for third-party auditors. In this article, we focus on the most common type, algorithm audits~\cite{birhane2024}, as they require less privileged access and are perceived as less invasive by the auditee. 

Casper et al.~\cite{casper2024} propose a typology for algorithm audits based on access level, distinguishing \textit{black-box}, \textit{grey-box}, \textit{white-box}\footnote{As defined by Casper et al.~\cite{casper2024}, black-box access only allows auditors to query a system and analyze its output, while white-box allows full access to a model, including feature weights and ability to fine-tune the model.  Grey-box audits are situated in-between, allowing auditors access to some of the inner workings of a model.}, and \textit{outside-the-box} access---the last of which involves accessing contextual information about a model such as methodology, documentation, and internal evaluations. They document important differences in audit flexibility between these types of access, and argue that audit effectiveness is dependent on the degree of access granted to auditors. The authors highlight that providing auditors with necessary access to systems is feasible, benefits the audited organization by establishing increased credibility and trustworthiness, and “allows for more meaningful oversight from audits”~\cite{casper2024}.\\

Auditors face important challenges in accessing sufficient high-quality data. Birhane et al. note that external auditors “typically struggle to access the information necessary to conduct a thorough investigation”~\cite{birhane2024}. Similarly, Raji et al. identify insufficient access as the main vulnerability to algorithm auditing, stating that “auditors must be granted access to enough data to conduct a robust review” but that it currently requires “extraordinary efforts for investigators to gain sufficient access for a thorough analysis”~\cite{raji2022}.

These risks have been identified as a primary issue in policy reports, including from the UK Centre for Data Ethics and Innovation 2021 report on bias in algorithmic decision-making~\cite{centrefordataethicsandinnovation2019}, the UK Information Commissioner's Office 2022 report on algorithm audits~\cite{informationcommissionersoffice2022}, and the AI Now Institute 2021 report on algorithmic accountability~\cite{adalovelaceinstitute2021}.

Although academic and regulatory work highlight the need for data access in algorithmic impact assessment~\cite{algorithmwatch2020}, the tools and regulations required to operationalize this oversight are lacking~\cite{ojewale2024}. This leaves auditors vulnerable to methodological skepticism and corporate retaliation. For instance, ProPublica's audit of COMPAS~\cite{angwin2016} received skeptic responses from the organization behind the model~\cite{angwin2016a} as well as from academics~\cite{flores2016} who dismissed statistical signals of discrimination. While these cases of skepticism have sparked some necessary discussions~\cite{chouldechova2017,corbett-davies2017,kleinberg2016}, such methodological disputes are in large part caused by low-access audits, and could therefore be avoided with a better approach to auditor access. In other cases, corporate pressure leads to the complete interruption of auditing efforts, such as the AlgorithmWatch investigation into Meta's Instagram algorithms~\cite{kayser-bril2021}.

\subsection{Quantitative audits for decision-making algorithms}
To audit black-box decision-making models without extensive access, researchers have proposed numerous methods using work-arounds such as training shadow models~\cite{tan2018} or testing for indirect influence of protected characteristics~\cite{adler2016}. As noted by Obermeyer et al., researchers often have to work ``from the outside'' and find workarounds to investigate algorithmic harms~\cite{obermeyer2019}. The prevalence of audit methodologies developed for low levels of access, including the absence of access to the model or the absence of labelled data, reflects the underlying issue: auditors lack appropriate access for meaningful oversight.

The value of quantitative approaches to measuring algorithmic discrimination has been widely discussed in the field~\cite{binns2021,mitchell2021}. Scholars have argued that focusing on mathematical formalization conflates the ML and policy problems~\cite{corbett-davies2023} and fails to address the underlying roots of discrimination~\cite{hoffmann2019}. However, even simple group fairness metrics can encourage clarifications around assumptions and definitions of `fairness', help shed light on underlying disparities and inconsistencies, and, by extension, inform policy~\cite{mitchell2021}. Quantifying disparities also has the potential to reveal issues that would otherwise go unnoticed, and be a first step towards addressing them. For instance, Angwin et al.'s audit of COMPAS first revealed racial disparities by comparing true and false positive rates between groups~\cite{angwin2016}; Obermeyer et al.'s audit of a healthcare algorithm similarly used calibration to demonstrate racial bias~\cite{obermeyer2019}; Jaiswal et al.~\cite{jaiswal2022} and Buolamwini and Gebru~\cite{buolamwini2018} used accuracy equity to evaluate facial recognition systems. In all these cases, group parity metrics were used as a starting point and diagnostic tool in auditors' assessments (also referred to as `harms discovery stage'~\cite{ojewale2024}). 

Group parity metrics help answer questions such as “do outcomes systematically differ between demographic groups?”. Castelnovo et al.~\cite{castelnovo2022} provide a comprehensive review and introduction to parity metrics. We share their perspective that the variety of existing metrics reflects fairness as a multi-faceted rather than absolute concept. These metrics have been shown to be efficient audit tools, provided that (i) the “right” measurement of discrimination is selected (which is context-dependent)~\cite{hellman2020}, (ii) the dataset used is appropriate, and (iii) the results are interpreted in-context. 
The question of appropriate metric selection and interpretation has received substantial scholarly attention~\cite{corbett-davies2017,kleinberg2016,ding2022,ruf2021}. Our work focuses on the second requirement, i.e. accurate estimations of parity metrics for independent auditors with limited data.

Limited data access can make group parity metrics harder to estimate. Besse et al.~\cite{besse2018} prescribe using confidence intervals rather than single values (which has been common practice) when computing parity metrics. Ji et al.~\cite{ji2020} have explored the question of the reliability of parity metrics estimations, focusing on cases where a relatively large dataset is available but only with a limited number of ground-truth labels. They propose a method to enhance datasets with this type of limitation, and observe the high impact of sample size on metric estimates. This effect is particularly damaging when the compared demographic groups are imbalanced. From this initial experiment, they conclude that “there can be high uncertainty in empirical estimates of groupwise fairness metrics, given the typical sizes of datasets used in machine learning”~\cite{ji2020}. 

In order to standardize auditing in the AI sector, important methodological challenges need to be addressed. The majority of proposed data collection or access methods in the algorithmic auditing literature are specific to user-facing systems (e.g. user audit\footnote{User audits collect data from the users themselves, e.g. through interviews or surveys.}, scraping\footnote{Scraping audits involve the direct collection of public-facing data on a platform, typically through an API.}, sock-puppets\footnote{Sock-puppet audits involve the creation of fake user accounts to collect data from a platform.} as presented by Sandvig et al.~\cite{sandvig2014}). User-engaged auditing has also gained the interest of practitioners and HCI researchers~\cite{shen2021,devos2022,deng2023}. Most audit studies in the literature are conducted on online platforms~\cite{bandy2021,ali2019} and interactive language models~\cite{mokander2023}, which have unique constraints and data collection methods. There are fewer examples of empirical audits for public sector decision-making algorithms. Therefore, practical considerations to align existing methodologies and metrics with the challenges of external data access are still a work in progress. 

Chen et al.~\cite{chen2021} note that due to limited access and the burden of proof being on users and auditors, proof of discrimination can only be gathered when “the problem becomes too extensive to ignore”. Without proper governance, audit-washing risks hindering progress towards rigorous algorithmic evaluations~\cite{informationcommissionersoffice2022,floridi2021,goodman2022,birhane2024}. Given the current scarcity of data access, a similar risk lies in organizations making low quality data available. This allows them to claim cooperativeness and transparency while effectively invisibilizing disparities enacted by their models~\cite{evans2023}. Robust third-party audits are necessary to counterbalance these risks.

\subsection{Protecting privacy in data sharing}
Computer science research in algorithmic fairness often assumes access to sensitive attributes (or `protected characteristics' in EU discrimination law terms), which may be reasonable when dealing with publicly available benchmark datasets, but less so in real world auditing contexts~\cite{veale2017,holstein2019}. Accessing and sharing such data is inherently constrained by the need to protect privacy~\cite{trask2024}, and poses a significant challenge in widening access to data while safeguarding individuals' privacy~\cite{stadler2022,belgodere2024}. In the EU, the Digital Services Act states that audited platforms should anonymize or pseudonymize personal data, unless doing so would render impossible the research purposes~\cite{europeanunion2022}. However, identifying privacy mechanisms that ensure compliance with data protection regulations and audit robustness remains an ongoing challenge. These technologies are deployed without sufficient evidence to guarantee their benefits~\cite{stadler2022,steed2022}. In turn, auditors are unable to assess the suitability of anonymized datasets to conduct meaningful algorithm audits.

In our work, we investigate two main types of privacy technologies, as reviewed by Gadotti et al.~\cite{gadotti2024a}: (i) record-level techniques---including pseudonymization and data minimization approaches---which reduce re-identification risks but do not fully guarantee private analysis, and (ii) aggregation techniques, including summary statistics or ML models approaches based upon `differential privacy' and `synthetic data'. While these privacy-enhancing technologies offer somewhat promising avenues for increased data access for auditors, their impact on third-party algorithm audits require further assessment to establish standardized auditing practices and ensure robust accountability.

\textit{Data minimization} as a principle involves limiting access, sharing only the necessary data for a given analysis, and aggregating when possible~\cite{cummings2024a}. For instance, de-identification (or pseudonymization) involves removing direct identifiers and falls within that category, despite being recognized as insufficient for public data sharing and sensitive data protection~\cite{gadotti2024a}. Still, data minimization is a stated goal in the EU's General Data Protection Regulation (GDPR), and can help address proprietary concerns for organizations. In specific cases, such as privileged access for auditors, data minimization can be sufficient, provided formalized audit standards for algorithms are established. However, researchers have warned that overemphasizing data minimization can limit access to important demographic information, therefore obscuring inequalities~\cite{galdonclavell2020,neftenov2023}.

\textit{Differential privacy} (DP) is an increasingly popular framework that provides formal privacy guarantees when sharing sensitive data, for both individual-level and aggregated data. It has gained popularity with use by LinkedIn~\cite{rogers2020}, Meta~\cite{nayak2020}, Apple~\cite{appledifferentialprivacyteam2017}, and the U.S. Census Bureau~\cite{populationreferencebureau2023}, to name a few. For algorithm auditing, DP can be achieved by adding a random amount of noise to aggregate statistics such as confusion matrices or parity metrics, allowing quantitative fairness assessments without releasing code or individual data. DP is known to work well for aggregate statistics that have low sensitivity, meaning that adding or removing an individual’s record does not significantly alter results~\cite{dwork2009}, as is the case with confusion matrices. DP can also be used to share individual-level data when combined with techniques such as synthetic data.  
However, while data released using differentially private mechanisms offer theoretical guarantees against re-identification, they could provide insufficient granularity or data quality for audit analysis, potentially affecting research outcomes~\cite{europeandigitalmediaobservatory2022}. Research by Imana, Korolova and Heidemann~\cite{imana2023} suggests that such privacy guarantees do not significantly hinder auditors' ability to achieve statistical confidence, provided the sample audience is increased and demographic groups are equally represented.
    
\textit{Synthetic data generation} aims to learn statistical properties of sensitive data in order to generate ``artificial'' data that are structurally and statistically similar to the original. This approach provides the auditor with a flexible alternative to access individual-level data, and is presented as a promising approach for private data sharing. As it attracts significant interest from practitioners~\cite{stadler2022,belgodere2024,sivizacaconde2024a}, it is crucial to evaluate its trustworthiness and suitability for algorithm audits. Synthetic data generation could limit audit reliability by introducing artifacts and random noise that reduce the overall quality of the data as well as remove statistical outliers~\cite{stadler2022,annamalai2024}, more likely to represent minorities. Pereira et al.~\cite{pereira2024} report that synthetic data has potential to be used for fairness evaluations, but find disparate reliability between generators. In practice, synthetic data is already being used for algorithm audits~\cite{braun2023} and presented as a viable tool in auditing frameworks, e.g. by consulting company Deloitte~\cite{eggers2019} and NGO Algorithm Audit~\cite{algorithmaudit}.

\begin{table*}[htb]
\centering
\begin{tabular}{lcccc}
\toprule
\multirow[b]{3}{*}{Access scenario} & \multicolumn{4}{c}{Auditor access} \\ \cmidrule{2-5}
 &
  \begin{tabular}[c]{@{}c@{}}Aggregate \\ statistics\end{tabular} &
  \begin{tabular}[c]{@{}c@{}}Individual-level\\data\end{tabular} &
  Model query &
  \begin{tabular}[c]{@{}c@{}}Model class \& parameters\\(excluding weights)\end{tabular} \\ \midrule
 \begin{tabular}[c]{@{}l@{}}(A) Group confusion \\ matrices only\end{tabular} & \cmark & \xmark & \xmark & \xmark \\
                           \begin{tabular}[c]{@{}l@{}}(B) Dataset with access \\ to model predictions\end{tabular} & \cmark & \cmark & \cmark & \xmark \\
                           \begin{tabular}[c]{@{}l@{}}(C) Dataset without access \\ to model predictions\end{tabular} & \cmark & \cmark & \xmark & \cmark \\ \bottomrule
\end{tabular}
\caption{Typology of third-party audit scenario and respective access to data, model, and parameters}
\label{tab:typology}
\end{table*}

\begin{table*}[htbp]
\centering
\begin{tabular}{lccccc}
\toprule
\multirow[l]{4.8}{*}{Access scenario} & \multicolumn{5}{c}{Data quality loss} \\ \cmidrule{2-6}
 &
  Subsampling &
  \begin{tabular}[c]{@{}c@{}}Missing\\features\end{tabular} &
  \begin{tabular}[c]{@{}c@{}}Missing\\values\end{tabular} &
  \begin{tabular}[c]{@{}c@{}}Differential Privacy\\aggregation\end{tabular} & 
  Synthetization \\ 
  \midrule
 \begin{tabular}[c]{@{}l@{}}(A) Group confusion\\matrices only\end{tabular} & \cmark & \xmark & \xmark & \cmark & \xmark \\
 \begin{tabular}[c]{@{}l@{}}(B) Dataset with access\\to model predictions\end{tabular} & \cmark & \cmark & \cmark & \xmark & \cmark \\
 \begin{tabular}[c]{@{}l@{}}(C) Dataset without access\\to model predictions\end{tabular} & \cmark & \cmark & \cmark & \xmark & \cmark \\ \bottomrule
 \addlinespace
\end{tabular}
\parbox{0.6\textwidth}{\caption{Mapping of access scenarios and experiments.\\\cmark indicate that the access scenario can be affected by this data quality loss.}}
\label{tab:exp-scenario}
\end{table*}

\section{Methods}

\subsection{Setting} \label{setting}
Our experiments assume the following standard setting: an independent actor (the auditor) accesses data to evaluate the parity of a decision-making algorithm developed or used by an organization (the auditee) with respect to a given characteristic (e.g. race, gender). The data may be received directly from the auditee or collected by other means, such as obtained from a third party. Compared to the training dataset originally used to develop the algorithm, the audit dataset can vary in its size, structure and quality. Through our experiments, we examine how changes in the audit data affect group-based parity metrics.

Table~\ref{tab:typology} summarizes the three modes of access for third-party auditors that we consider. For each access scenario, we evaluate the impact of data quality and minimization factors on metric reliability.

In \textbf{Access Scenario A}, auditors are provided with confusion matrices representing the model outputs for the requested demographic groups. Access is restricted to these aggregate statistics, which auditors use to calculate their selected metrics.

\textit{Example.} In 2021, the Netherlands Institute for Human Rights investigated the Tax Administration’s Childcare Allowance fraud detection algorithm. Their analysis was solely based on confusion matrices to compare past algorithm decisions across gender groups. Auditors did not have access to predictor data and could not query the model. Despite this limited access, the statistical assessment served in legal procedures against the government as evidence of indirect discrimination~\cite{hoogenboom2022}.

In \textbf{Access Scenario B}, auditors access individual-level data, containing both the ground-truth and the demographic groups of interest. The dataset may already contain predictions, or the auditors may query the model to obtain them (e.g. through API access, typically with a limited number of queries), and compute metrics.

\textit{Examples.} In 2016, ProPublica’s audit of COMPAS~\cite{angwin2016} used individual-level data obtained through a public records request, which contained model outputs and compiled ground truths, to assess racial bias. 
In 2021, Koulish and Evans’ audit of the Immigration and Customs Enforcement Risk Classification Assessment algorithm~\cite{koulish2021} had a similar access setup, with labelled individual-level data obtained through court orders. They were able to assess disparities based on various social vulnerability factors.
In 2016, Lum  and Isaac’s audit of PredPol~\cite{lum2016} used de-identified police records and a synthetic population dataset to query the model and compare outputs on racial bias. Rather than receiving a labelled dataset, auditors had access to data and to the model separately.

In \textbf{Access Scenario C}, auditors access individual-level data containing ground-truth and demographic groups. They do not have predictions and no direct or indirect model access (i.e. uncooperative auditee), but have knowledge of the model class and its parameters. Auditors replicate the audited model by training on a portion of the dataset and compute metrics on the remaining subset. This access scenario can arise in data donation settings, whereby auditors have data but no access to the model or its outputs. Provided sufficient contextual information about the model, they can attempt to replicate the model.

\textit{Example.} In 2023, non-profit \textit{La Quadrature du Net} audited the French family allowance fund's fraud detection algorithm with a comparable setting~\cite{laquadraturedunet2023}. With only access to the objective function of the model and no query access or labelled dataset, auditors replicated the model and assessed it with simulated archetype data.\\

For all three scenarios, our study makes several important assumptions. First, we focus on binary classifiers as the target of our simulated audits, as they are commonly used in (social) prediction tasks. We address how our results may extend to multi-classifiers and risk scorers—both increasingly common as well—in the Discussion section. In Scenarios A and B, the audit is a ``black-box'' level of access, whereby the auditor does not have direct, transparent access to the system, and can only observe a sample of inputs and outputs. We found this to be the most common configuration in empirical audits of public sector algorithms, with only very few cases of privileged “white-box” access (e.g.~\cite{seidelin2022}). Third, the audit data includes the ground-truth against which to compare the model prediction. This requires a formal definition for said ground-truth, which is what many group parity metrics rely on. Finally, the audit data includes the protected characteristics for which auditors wish to run comparisons on. Past research by Kallus et al.~\cite{kallus2022} demonstrates that relying on proxies for demographic labels is unreliable for disparity assessments. Including demographic labels in the shared dataset implies a need for higher privacy safeguards.

\subsection{Experiments}
We examine five key risks of quality loss faced by auditors which may undermine the accuracy and reliability of their assessment: (1) low sample size, (2) missing predictors, (3) disparate rates of missing data across groups, (4) low privacy budgets in differential privacy, and (5) low utility synthetic data. We conduct five distinct experiments to assess their impact on metric estimation. The rationale and methodology is described below.

\subsubsection{Reduction of sample size}
Organizations or data curators share samples of various sizes, which may not be informed by the auditor’s needs. The size of the audit dataset needs to satisfy both data minimization and audit reliability requirements. In this experiment, to test how sample size impacts audit reliability, the auditor receives a 1 to 100\% subset of the audit dataset.\\
\textit{Affected scenarios:} A, B and C.

\subsubsection{Removal of features} 
Features can be removed by an organization for data minimization purposes, or be missing due to differences in data curation across organizations. For instance, if the sample is collected through a third party rather than received from the audited organization, all model predictors may not be included. This has been hypothesized to be the case with the audit dataset used by ProPublica for its assessment of COMPAS~\cite{tan2018}, which may have altered the parity metrics. To test how feature removal impacts audit reliability, we gradually remove features from the audit dataset before obtaining predictions from the model and running the audit. The features are ordered by importance, from the weakest to the strongest predictor according to Shapley values~\cite{lundberg2017}.\\
\textit{Affected access scenarios:} B and C. In Scenario B, if the data curator shares a dataset including model predictions, from which some features are removed as part of a data minimization process, then the missing features do not affect metrics. However, if the predictions are obtained separately by auditors, or if the missing features are involuntary on the data curator’s part, then predictions and metric may be affected. Similarly, for Scenario C, the replicated model may be missing important predictors, which would skew metric values despite having the right parameters for model training. 

\subsubsection{Disparate incompleteness}
In various domains, underprivileged groups tend to have higher rates of missing data, which is one mechanism through which a model can become biased~\cite{getzen2023,teeple2023}. For example, Wilson et al.~\cite{wilson2021} audited an automated candidate screening system and found that distributions of missing data were significantly different across demographic groups. Zhang and Long~\cite{zhang2021} conducted experiments measuring fairness criteria on samples with missing data, and found that bias in fairness estimations is larger in the presence of missing values. To test how disparate incompleteness impacts audit reliability, we randomly remove 1-60\% of values from the model’s top five features, for the underprivileged group only. This allows us to verify whether non-randomly distributed missing data (i.e. where the underprivileged groups have higher rates of missing data as compared to privileged groups) in the audit dataset affect parity metrics estimations.\\
\textit{Affected access scenarios:} similar to above, B and C.

\subsubsection{Differential Privacy (DP)}
Differentially-private mechanisms that inject random noise in features have been proposed to reduce re-identification risks. Research has shown that not all aggregate statistics can be accurately shared using differential privacy~\cite{houssiau2022}. Indeed, random noise may skew aggregate statistics and limit the reliability of parity metrics. To test if differentially private statistics can allow for both privacy protection and metric reliability, we compare parity metrics obtained using confusion matrices with an $\epsilon$-DP mechanism, by varying the privacy budget $\epsilon$. We use Laplace mechanism~\cite{steed2022,dwork2006a} with a sensitivity of 1 and privacy budget ranging from $\epsilon=0.01$ (very strong privacy protection) to $\epsilon=10$ (less restrictive privacy protection), reflecting the range typically used in DP research~\cite{hsu2014}.\\
\textit{Affected access scenario:} A.

\subsubsection{Synthetic Data Generation}
Synthetic data has been presented as a solution to ensure privacy while maintaining high utility and flexibility. To test how generation mechanisms impact metric reliability, we generate synthetic data using several statistical and machine learning generation models: the Gaussian Copula Synthesizer, CT-GAN Synthesizer, and Copula-GAN Synthesizer from the Synthetic Data Vault library~\cite{patki2016}, as well as PrivBayes~\cite{zhang2017} using the DPART library. For each model and dataset, we generate 100 synthetic audit samples. We use random samples ($n=7,500$ for the NIJ Recidivism dataset and $n=20,000$ for the ACS Public Coverage dataset) from the real audit dataset to train and generate synthetic samples of the same size.\\
\textit{Affected access scenarios:} B and C.


\subsection{Research Design}

\begin{figure*}[!htbp]
\centering
\includegraphics*[width=0.9\linewidth]{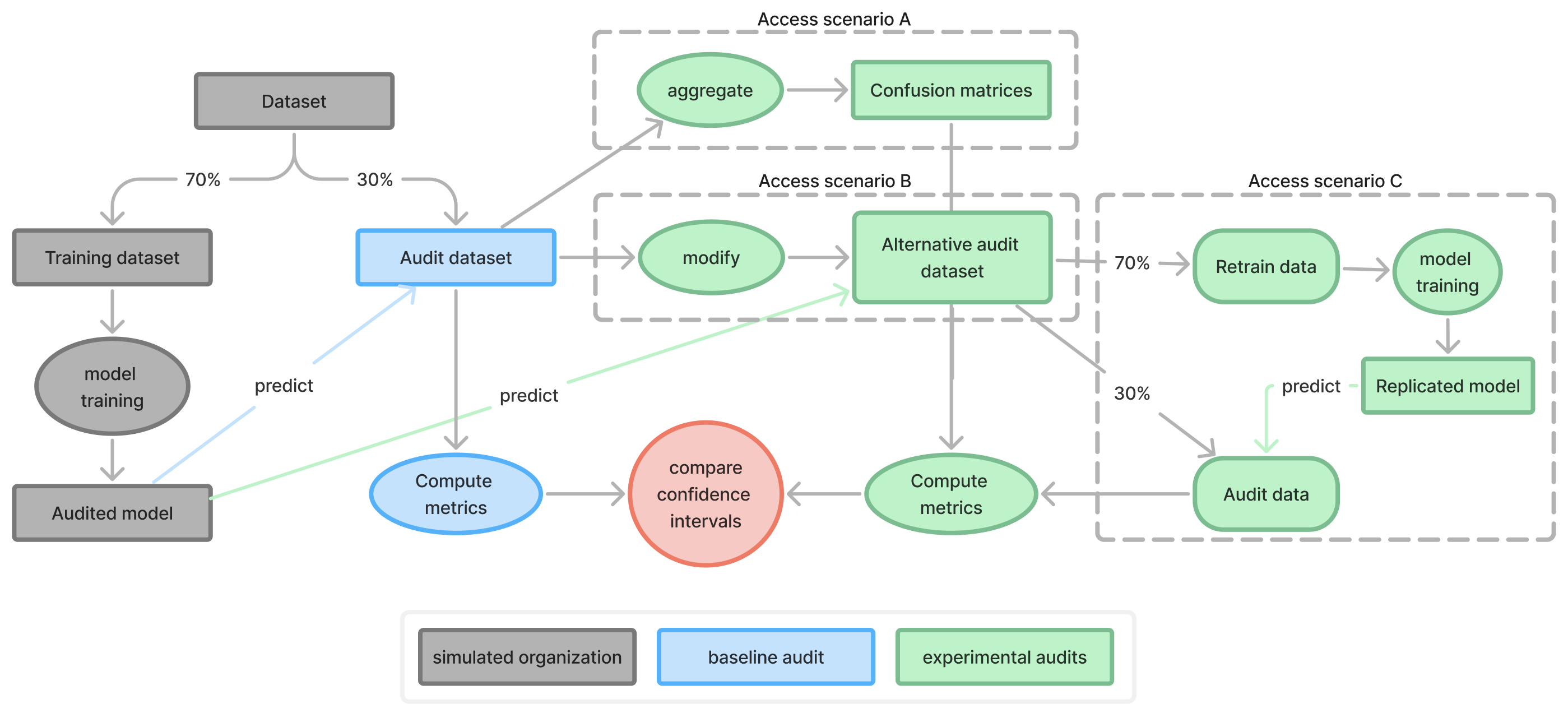}
\caption{Experimental design flowchart, distinguishing between the simulated organization (in grey), baseline audit (in blue), and audits under Access Scenarios A, B and C (in green).}
\label{fig:flowchart}
\Description[A flowchart for the experimental design used in the study.]{The flowchart separates actions conducted by the simulated organization from the baseline, ideal conditions audit, and from the experimental audits conducted on lower quality data. The simulated organization has a dataset, from which we take 70\% to train the model; the other 30\% is the audit dataset. We obtain predictions from the model for the audit dataset, and compute metrics with bootstrapping to obtain baseline 95\% confidence intervals. For experimental audits, the dataset is modified before obtaining model predictions and computing metrics. The baseline and experimental confidence intervals are then compared.}
\end{figure*}

Figure~\ref{fig:flowchart} shows our experimental design. First, we split the dataset between training (70\%) and auditing (30\%) sets. In order to control for sample variability, we repeat experiments over 100 random train-audit splits. The training set is used by the simulated organization to train, tune, and validate its classification model. Second, predictions are obtained for the auditing set from the audited model, and group parity metrics are computed with bootstrapping (500 repetitions). We obtain 95\% confidence intervals for each metric~\cite{besse2018}. These 95\% confidence intervals are used as baselines and proxies for the metrics’ ground truth. This constitutes the \textit{baseline audit}, where the audit set is in its optimal state.

The audit set is then modified to simulate an audit with lower data quality using experiments (1) to (5). For each experiment, these results are compared with the baseline audit results, and we observe whether and to what extent the two confidence intervals overlap. Note that, for Access Scenario C, the audit set is split again, with 70\% being used by auditors to re-train the model, and the remaining 30\% used to compute metrics.

\subsection{Measuring audit reliability}

The baseline and experimental 95\% confidence intervals are compared on two complementary criteria. We observe whether the two intervals have the same \textit{configuration} to determine whether the audit under said conditions results in:
\begin{itemize}[noitemsep, parsep=2pt, topsep=3pt, font=\normalfont\itshape, labelindent=\parindent]
    \item \textit{Accurate interpretation} if the configurations of the baseline and experiment intervals match;
    \item \textit{Type 1 error} if the auditor identifies a disparity affecting group A or B, when in reality there is none;
    \item \textit{Type 2 error} if the auditor identifies there is no disparity, when in fact there is;
    \item \textit{Reverse error} if e.g., the auditor identifies a disparity affecting group A when in reality it affects group B.
\end{itemize} 

We also measure metric \textit{reliability} by computing the proportion of values within the experimental results that overlap with the baseline interval. A low proportion indicates that auditing under these given conditions leads to unreliable results.

\subsection{Case studies}

\subsubsection{Datasets}
We build our case studies from two publicly available datasets. 

\begin{table*}[!ht]
\centering
\begin{tabular}{llllllll}
\toprule
Domain & Dataset & Prediction task & Data origin &
\begin{tabular}[c]{@{}l@{}}Sample\\size\end{tabular} & Model class & \begin{tabular}[c]{@{}l@{}}Performance\\metrics\end{tabular} \\ 
\midrule
\begin{tabular}[c]{@{}l@{}}Criminal\\justice\end{tabular} &
  \begin{tabular}[c]{@{}l@{}}NIJ\\Recidivism\end{tabular} &
  \begin{tabular}[c]{@{}l@{}}Recidivism\\within 3 years\\of parole\\release\end{tabular} &
  \begin{tabular}[c]{@{}l@{}}State of Georgia, US\\ (2013-2015)\end{tabular} &
  25,835 &
  XGBoost &
  \begin{tabular}[c]{@{}l@{}}Accuracy = 0.68\\ Balanced acc. = 0.66\\ Precision = 0.68\\ Recall = 0.83\\ F1 Score = 0.75\\ ROC-AUC = 0.73\end{tabular} \\ \cmidrule(l){1-7}
Healthcare &
  \begin{tabular}[c]{@{}l@{}}ACS Public \\ Coverage\end{tabular} &
  \begin{tabular}[c]{@{}l@{}}Public health\\insurance\\coverage\end{tabular} &
  \begin{tabular}[c]{@{}l@{}}State of NY, US \\ (2015-2018)\end{tabular} &
  221,702 &
  XGBoost &
  \begin{tabular}[c]{@{}l@{}}Accuracy = 0.77\\ Balanced acc. = 0.72\\ Precision =  0.74\\ Recall =  0.57\\ F1 Score = 0.64\\ ROC-AUC = 0.81\end{tabular} \\
  \bottomrule
  \addlinespace
\end{tabular}
\caption{Description of case studies}
\label{tab:datasets}
\end{table*}

\paragraph{NIJ Recidivism dataset.} We selected the United States National Institute of Justice Recidivism Forecasting Challenge dataset, a publicly available dataset that fits a relevant use case for public sector algorithm auditing, i.e. recidivism prediction to inform decisions on pretrial and probation release. The dataset comprises over 25,000 individuals released from prison on parole between 2013 and 2015 in the State of Georgia. We assume that the auditee organization builds a model predicting recidivism within 3 years of release, and the auditor investigates racial disparities, with race defined as a Black/White binary (the only two categories provided in the dataset). The dataset has low class imbalance with 14,904 recidivists divided into 8,713 Black and 6,191 White, and 10,931 non-recidivists divided into 6,134 Black and 4,797 White. 

\paragraph{ACS Public Coverage.} We selected the American Community Survey (ACS) Public Use Microdata Sample (PUMS), accessed using Folktables~\cite{ding2022}. This dataset is publicly available and was created for the empirical study of algorithmic fairness. We assume that the auditee organization follows the prediction task defined by the authors, predicting whether an individual is covered by public health insurance, with the policy goal of lowering healthcare access inequality. Predictive models are increasingly used in healthcare to target this type of interventions (e.g.~\cite{obermeyer2019,agencyforhealthcareresearchandquality2014,strika2024}). In our simulation, a public agency uses the model to target interventions and allocate resources to reduce healthcare coverage gaps among vulnerable populations (in this case, low-income individuals). The population is US low-income individuals under 65, non-eligible for Medicare. The dataset consists of 19 features. We train and audit the model on data from the state of New York between 2015 and 2018. We filter the sample, keeping only individuals identifying as “White alone” or “Black or Native American alone”, and audit the model for racial bias. The dataset has high class imbalance on both race groups and the target variable, with 176,187 White individuals versus 45,515 Black/Natives, and 139,340 non-covered individuals versus 82,362 covered individuals.

\subsubsection{Models training} For each dataset, we trained XGBoost, Random Forest, Histogram-based Gradient Boosting Classification Tree, and Logistic Regression models. All models achieved an accuracy above 60\%. We performed experiments on each model to test whether the model class impacts results, which we found not to be the case. Therefore, we hereafter focus on XGBoost models as they performed best on both datasets. Performance metrics are reported in Table~\ref{tab:datasets}.

We also trained a Differentially Private Random Forest (DP-RF) model~\cite{holohan2019} for each dataset, with privacy budgets of $\epsilon=0.1$, $\epsilon=1$, and $\epsilon=10$. Compared to their non-DP Random Forest counterparts, the DP models show a reduction in accuracy from 67\% to 62-64\% (depending on the $\epsilon$ parameter) on the NIJ dataset. \Cref{results_dprf} compares audit reliability between DP and non-DP models for experiments (1) to (3), under Access Scenario B.

\subsubsection{Low and high disparity case studies} On both Recidivism and Public Coverage prediction tasks, the trained models exhibit relatively low racial disparity. To test metric reliability in high disparity settings, we create alternate cases where the model exhibits high racial disparity. In order to skew the measured value of parity metrics, we reassign 95\% of those who receive a positive prediction to the underprivileged group, and reassign the rest to the privileged group.

\subsection{Disparity Metrics}
In order to test for unethical or unlawful disparities between groups, auditors and organizations use different metrics based on the context and application of the model. To reflect this variability, we performed our experiments on a range of commonly-used group parity metrics. We found that results generalized across these metrics, and therefore include figures for a single relevant metric for each case study in \Cref{sec:results}, for conciseness purposes. We refer the reader to \Cref{sec:appendix} for figures that include several metrics, which demonstrate that reliability patterns are similar across group parity metrics. 

For the recidivism case study, we focus on \textit{Statistical Parity Difference} (SPD), also called \textit{Demographic Parity}, which captures whether each group receives positive predictions at an equal rate, regardless of the true outcomes. We select this metric because auditors would likely monitor the probability of being predicted as a future recidivist without accounting for ground-truth labels, given the historical racial bias associated with police re-arrests~\cite{heaven2020}.

For the public health case study, we focus on \textit{Average Odds Difference} (AOD), which measures the average difference in false positive rates and true positive rates between privileged and unprivileged groups. Auditors may select this metric to account for false positives, which lead to a negative outcome (i.e. limited access to resources and missing out on policies aimed at reducing coverage gaps), as compared to true positives which represent the neutral outcome (i.e. individuals who have public coverage and do not receive misallocated resources).

All differences are computed as underprivileged minus privileged groups, with 0 indicating parity. As an example of interpretation, for the recidivism case study, a negative \textit{SPD} value would indicate that Black defendants (the underprivileged group) are more likely to be labelled as future recidivists, and hence to receive a negative outcome (refused parole release). For the public coverage case study, a negative \textit{AOD} value would indicate that Black individuals are more likely to be incorrectly predicted as public coverage beneficiaries, thereby disproportionately missing out on much needed support.

\section{Results}
\label{sec:results}

\subsection{Access Scenario A (Aggregated data)} 

\subsubsection{Differential Privacy and sample size}
\begin{figure*}[!htbp]
\centering
\includegraphics*[width=0.85\linewidth]{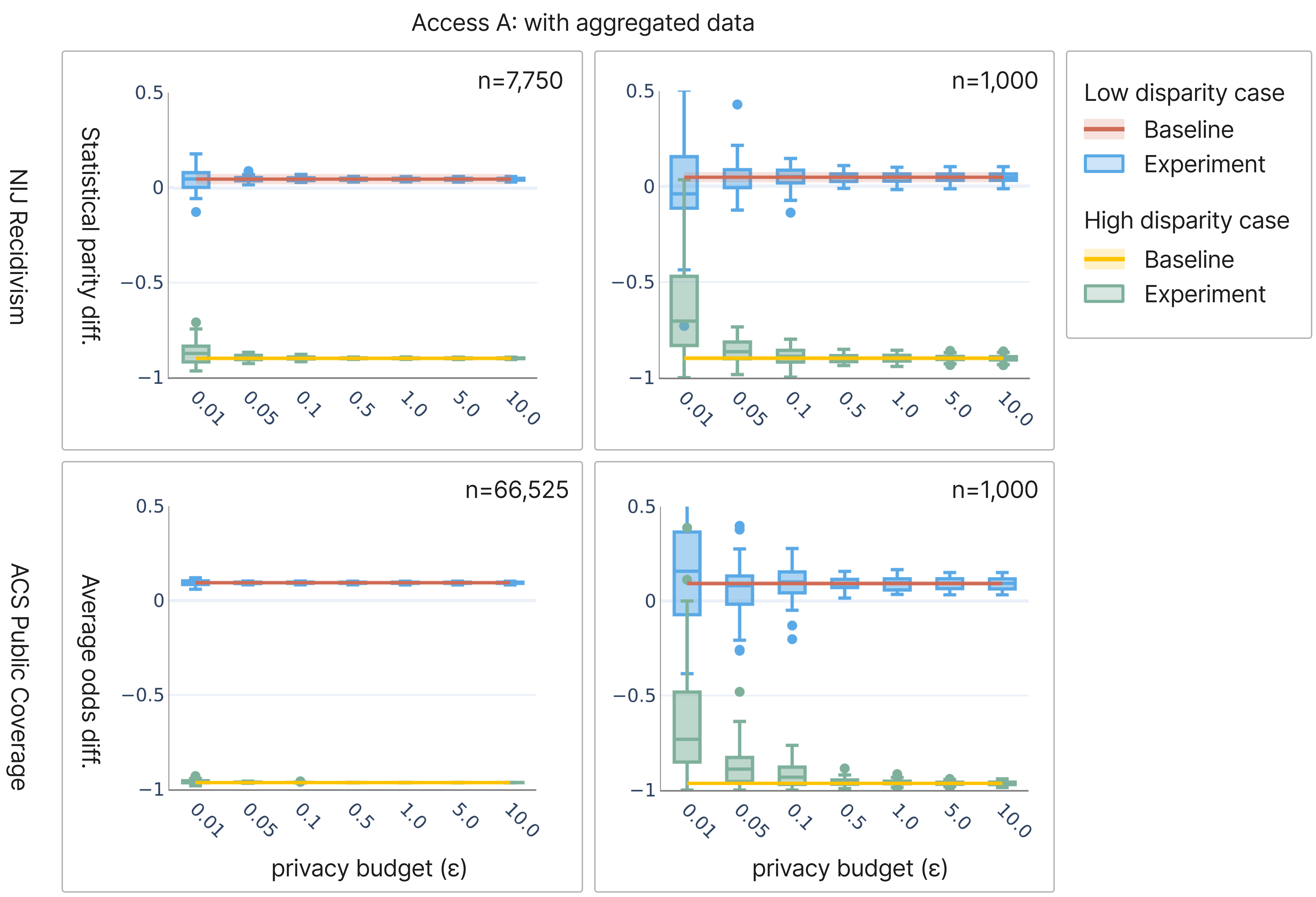}
\caption{\textbf{Effect of differential privacy on metric reliability} with full sample size and with a reduced ($n=1,000$) sample size.\\
The red and yellow lines represent median values for the baseline audit, while the colored areas represent 95\% confidence intervals obtained through bootstrapping and repetitions over dataset splits. Box plots represent metric estimations obtained from the DP aggregates at given privacy budgets, with bootstrapping and repetitions over dataset splits. In the box plots, middle lines represent medians.}
\label{fig:exp4}
\Description[Box plots for metric estimations across epsilon values]{Visualizations representing (a) confidence intervals for baseline audits and (b) box plots for experiment values, across a range of epsilon values (from 0.01 to 10), for both low and high disparity cases. The figure includes 4 subplots: one for the NIJ dataset at full audit sample size ($n=7,750$), one for the NIJ dataset at $n=1,000$, one for the ACS dataset at full audit sample size ($n=66,525$), and one for the ACS dataset at $n=1,000$. In all cases, lowering the privacy budget leads to broader ranges of metric estimations. This effect is significantly stronger in the $n=1,000$ plots, where very low privacy budgets lead to completely unreliable estimates.}
\end{figure*}

\Cref{fig:exp4} shows that differentially private confusion matrices with added random noise are generally highly accurate across privacy budget $\epsilon$. Under Access Scenario A, metrics only potentially become unreliable when the $\epsilon$ parameter is set at very low values ($\epsilon<0.05$), which are unlikely to be used in practice, or when the sample is small ($n<5,000$). For instance, with a sample of $n=1,000$, we find the minimal privacy budget to obtain reliable estimations to be around $\epsilon=0.5$ (see ~\Cref{fig:exp4}), while a sample of $n>5,000$ can yield reliable estimations with a lower privacy budget of $\epsilon=0.05$. ~\Cref{tab:accessA-sum} summarizes the overlap between estimations and their corresponding baseline. While some privacy budgets yield relatively low overlap values, \Cref{fig:exp4} shows that even with small samples (for $n=1000$), a privacy budget of $\epsilon$=0.5 or above yield estimations that are highly unlikely to lead to interpretation errors for auditors.  Therefore, differentially private statistics can simultaneously achieve audit reliability and strong privacy protection for individuals whose data is included in the audit dataset.

\begin{table*}[!htbp]
  \centering
  \begin{tabular}{lllll >{\centering\arraybackslash}m{1.1cm} >{\centering\arraybackslash}m{1.1cm} >{\centering\arraybackslash}m{1.1cm}}
    \toprule
    & & & & \multicolumn{3}{c}{\begin{tabular}[c]{@{}c@{}}Proportion of values\\within the baseline 95\% CI\end{tabular}} \\ \cmidrule(l){5-7}
    Experiment & Dataset & Metric & Disparity level & $\epsilon$ = 0.05 & $\epsilon$ = 0.5 & $\epsilon$ = 1 \\ \midrule
    
    \multirow{4}{*}{\begin{tabular}[c]{@{}l@{}}Differential Privacy\\with full sample size\end{tabular}} 
    & \multirow{2}{*}{\begin{tabular}[c]{@{}l@{}}ACS ($n=66,525$)\end{tabular}} 
    &  \multirow{2}{*}{AOD} & High & {\color[HTML]{34A853}0.73} & {\color[HTML]{34A853}1} & {\color[HTML]{34A853}1} \\
    & & & Low & {\color[HTML]{34A853}0.99} & {\color[HTML]{34A853}1} & {\color[HTML]{34A853}1} \\
    \addlinespace
    & \multirow{2}{*}{\begin{tabular}[c]{@{}l@{}}NIJ ($n=7,750$)\end{tabular}} & \multirow{2}{*}{SPD} & High & 0.63 & {\color[HTML]{34A853}1} & {\color[HTML]{34A853}1} \\
    & & & Low & {\color[HTML]{34A853}0.94} & {\color[HTML]{34A853}1} & {\color[HTML]{34A853}1} \\ 
    \cmidrule(l){2-7}
    
    \multirow{4}{*}{\begin{tabular}[c]{@{}l@{}}Differential Privacy\\with $n=5,000$\end{tabular}}
    & \multirow{2}{*}{ACS} 
    & \multirow{2}{*}{AOD} & High & {\color[HTML]{EA4335}0.10} & 0.44 & 0.46 \\
    & & & Low & {\color[HTML]{EA4335}0.06} & {\color[HTML]{EA4335}0.26} & 0.32 \\
    \addlinespace
    & \multirow{2}{*}{NIJ}
    & \multirow{2}{*}{SPD} & High & 0.45 & {\color[HTML]{34A853}0.97} & {\color[HTML]{34A853}0.99} \\
    & & & Low &  {\color[HTML]{EA4335}0.28} & 0.63 & 0.65 \\
    \cmidrule(l){2-7}

    \multirow{4}{*}{\begin{tabular}[c]{@{}l@{}}Differential Privacy\\with $n=1,000$\end{tabular}}
    & \multirow{2}{*}{ACS}
    & \multirow{2}{*}{AOD} & High & {\color[HTML]{EA4335}0.04} & {\color[HTML]{EA4335}0.08} & {\color[HTML]{EA4335}0.20} \\
    & & & Low & {\color[HTML]{EA4335}0.06} & {\color[HTML]{EA4335}0.26} & 0.32 \\
    \addlinespace
    & \multirow{2}{*}{NIJ}
    & \multirow{2}{*}{SPD} & High & {\color[HTML]{EA4335}0.12} & 0.41 & 0.52 \\
    & & & Low & {\color[HTML]{EA4335}0.28} & 0.63 & 0.65 \\
    
    \bottomrule
    \addlinespace
  \end{tabular}
  \caption{Proportion of metric values within the baseline 95\% confidence interval based on the $\epsilon$ parameter and sample size.\\Values $\geq$ 0.70 (indicating high overlap between estimations and the baseline) are indicated in green, and values $\leq$ 0.30 (indicating low overlap) are indicated in red.}
  \label{tab:accessA-sum}
\end{table*}

\subsection{Access Scenarios B and C (Individual-level data)}

\subsubsection{Reduction of sample size}
\begin{figure*}[!htbp]
    \centering
    \includegraphics*[width=0.855\linewidth]{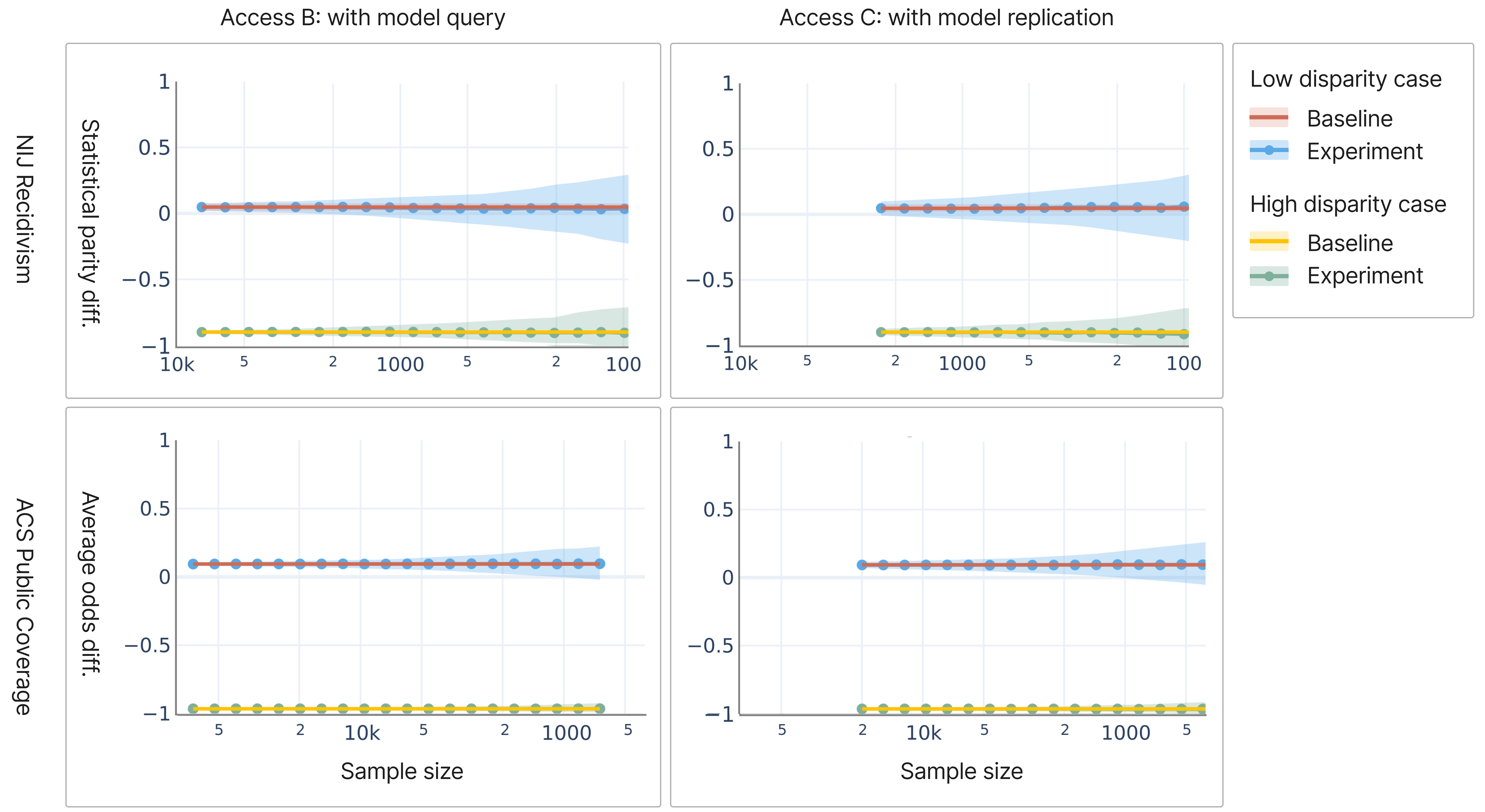}
    \caption{Effect of sample size on metric reliability for Access B (left) and Access C (right). The red and yellow lines represent median values for baselines. The blue and green dots represent median values for experiments, from 100\% to 1\% of the available sample in each case. Note that there are less data points on the Access C plots, as we only have 30\% of the full audit dataset available under this scenario (the other 70\% being used to retrain the model).}
    \label{fig:exp1}
    \Description[Line plots for metric estimations across audit sample size]{The figure includes 2 subplots for each dataset; one for Access Scenario B (with model predictions) and one for Access Scenario C (with model replication). Each plot displays baseline confidence intervals and experiment intervals across sample sizes, for both low and high disparity cases. In all cases, the experiment intervals become broader as the sample size decreases. At equal sample sizes, the confidence interval ranges are similar between access B and C.}
\end{figure*}

\Cref{fig:exp1} shows that, when reducing audit sample sizes, median values for metric estimations remain within the baseline confidence interval. However, the range significantly widens for both datasets under $n=1,000$, indicating low audit reliability with smaller samples. On average, reducing the sample size by 70\% reduces the proportion of values within the baseline confidence interval from 100\% to 68\% (\Cref{tab:accessBC-sum}). This variability is problematic when the metrics in fact indicate parity, as it can lead to Type 1 errors. In low disparity cases, it can also lead to Type 2 errors, or ‘reverse’ errors (finding a signal of bias towards the wrong group). For instance, for a \textit{SPD} ground truth of $-0.07$, over 10\% of samples at $n=1,000$ (or 25\% of samples at $n=500$) yield positive values and mislead auditors’ interpretations. For the recidivism case study, this means auditors could falsely interpret that the model slightly disadvantages White defendants, when it actually disadvantages Black defendants. Such interpretation errors are highly unlikely for highly biased models, but metric estimations can still significantly over or under-estimate group disparities with smaller samples. 

As shown in Figure~\ref{fig:exp1} and Table~\ref{tab:accessBC-sum}, audit reliability in Access Scenario C is equivalent to that in Access Scenario B at equal audit sample size. However, for the recidivism case study, the sample size would need to be increased by approximately 160\% compared to the baseline in order to achieve an average overlap of at least 90\% with the baseline (while maintaining a 70\%-30\% (re)train-audit split).

\begin{figure*}[!htbp]
    \centering
    \includegraphics*[width=0.855\linewidth]{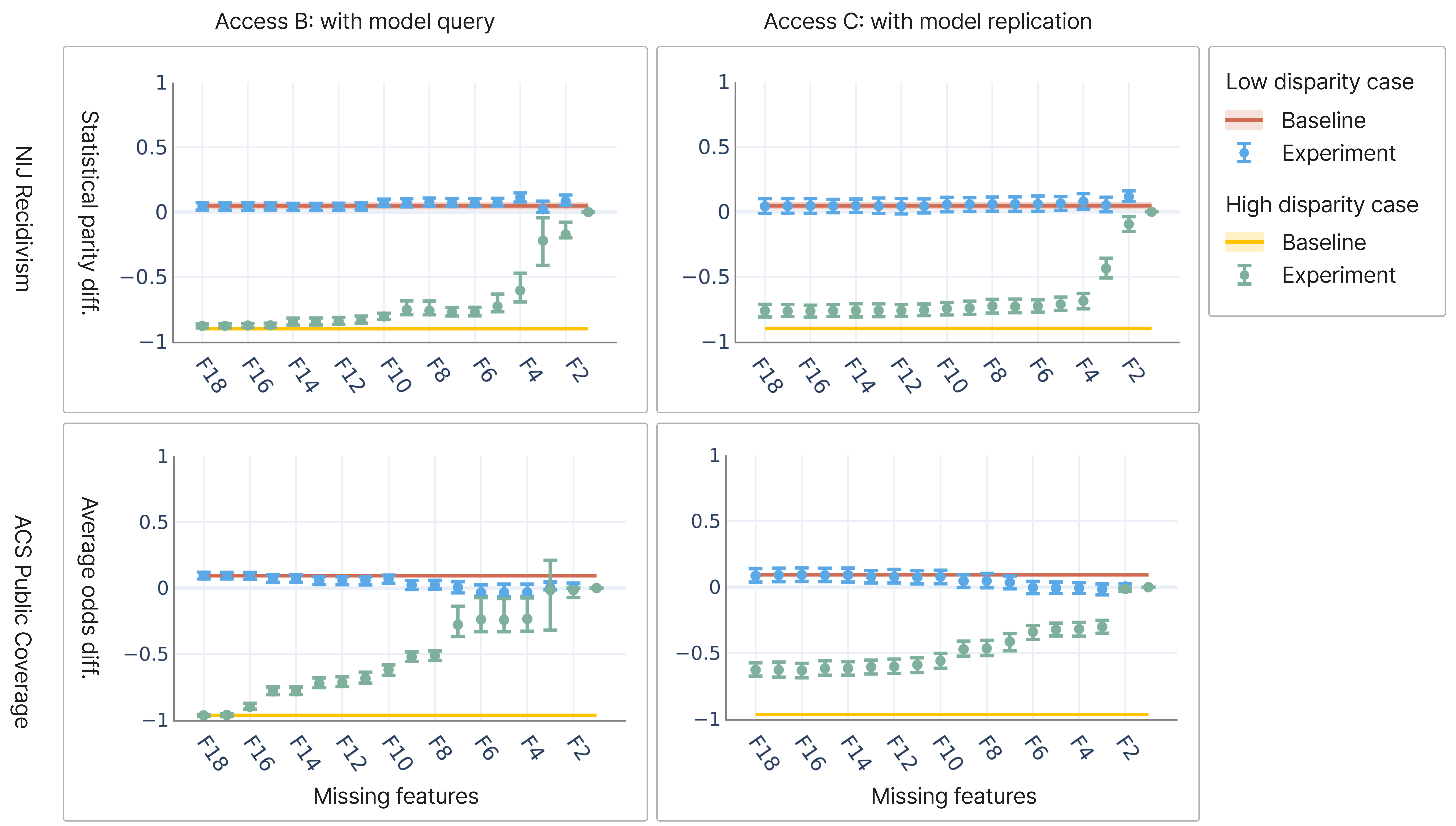}
    \caption{Effect of missing features on metric reliability for Access B (left) and Access C (right). On the x-axis, features are ordered by increasing order of importance. Plots are cumulative (e.g. at the F14 point, features 14 to 18 are missing from the audit dataset). The red and yellow lines represent median values for baselines, while the blue and green dots represent median values for experiments. The error bounds represent 95\% confidence intervals obtained through bootstrapping and repetitions over dataset splits.}
    \label{fig:exp2}
    \Description[Scatter plots for metric estimations based on feature removal]{The figure includes 2 subplots for each dataset; one for Access Scenario B (with model predictions) and one for Access Scenario C (with model replication). Each plot displays baseline confidence intervals and experiment intervals across sample sizes, for both low and high disparity cases. For high disparity cases, the experiment values deviate towards 0 as the cumulative number of missing features increases.}
\end{figure*}

\begin{figure*}[!htbp]
\centering
\includegraphics*[width=0.92\linewidth]{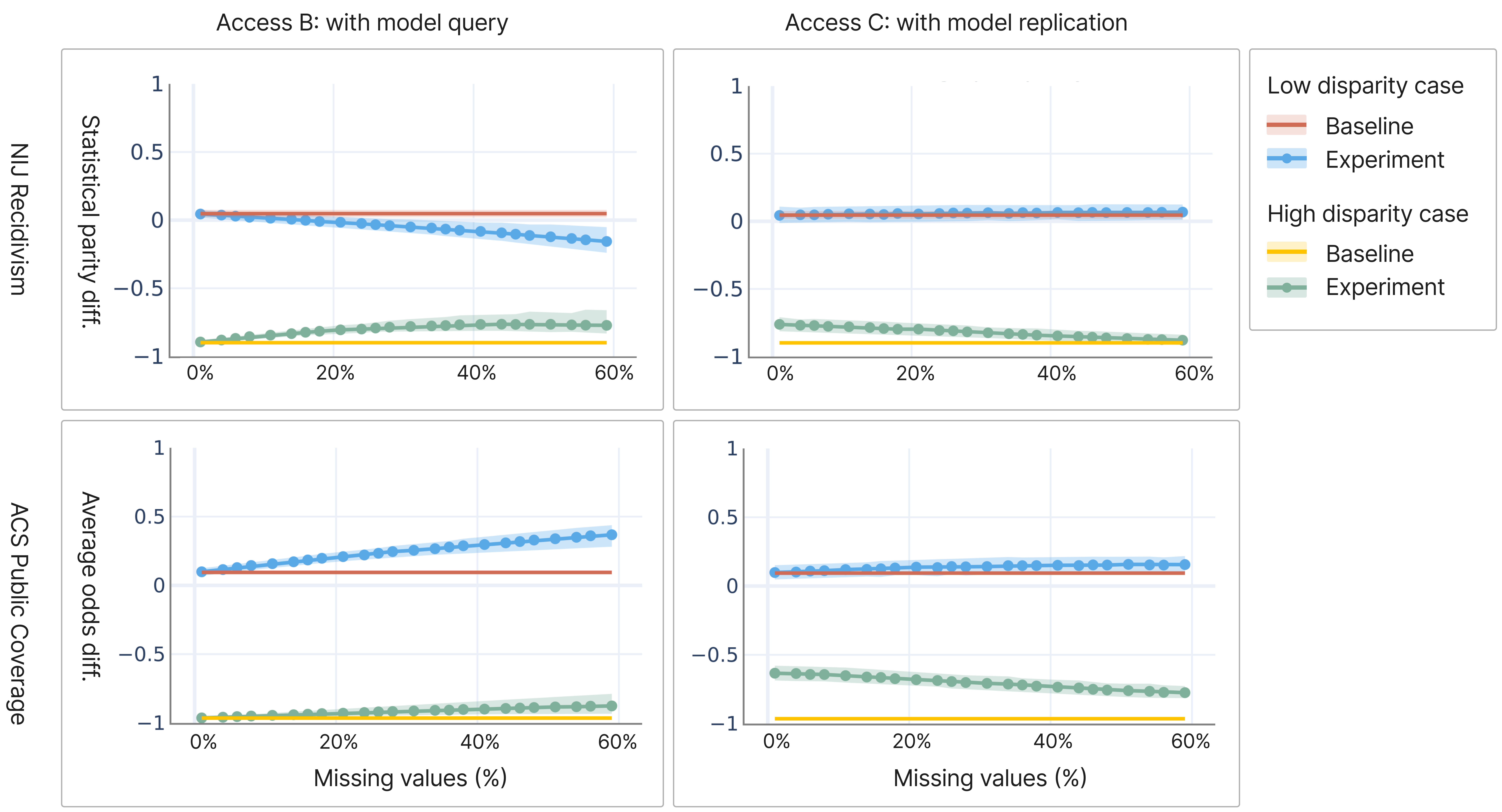}
\caption{\textbf{Effect of disparate missing values rates on metric reliability} for Access B (left) and Access C (right). The red and yellow lines represent median values for baselines, while the blue and green dots represent median values for experiments. The colored areas represent 95\% confidence intervals obtained through bootstrapping and repetitions over dataset splits.}
\label{fig:exp3}
\Description[Scatter plots for metric estimations based on rates of missing values among the underprivileged group]{The figure includes 2 subplots for each dataset; one for Access Scenario B (with model predictions) and one for Access Scenario C (with model replication). Each plot displays baseline confidence intervals and experiment intervals across sample sizes, for both low and high racial disparity cases. For access scenario B, experiment intervals deviate from the baseline as the rate of missing values increases. For access scenario C, experiment intervals remain close to the baseline.}
\end{figure*}

\begin{figure*}[!htbp]
\centering
\includegraphics*[width=0.92\linewidth]{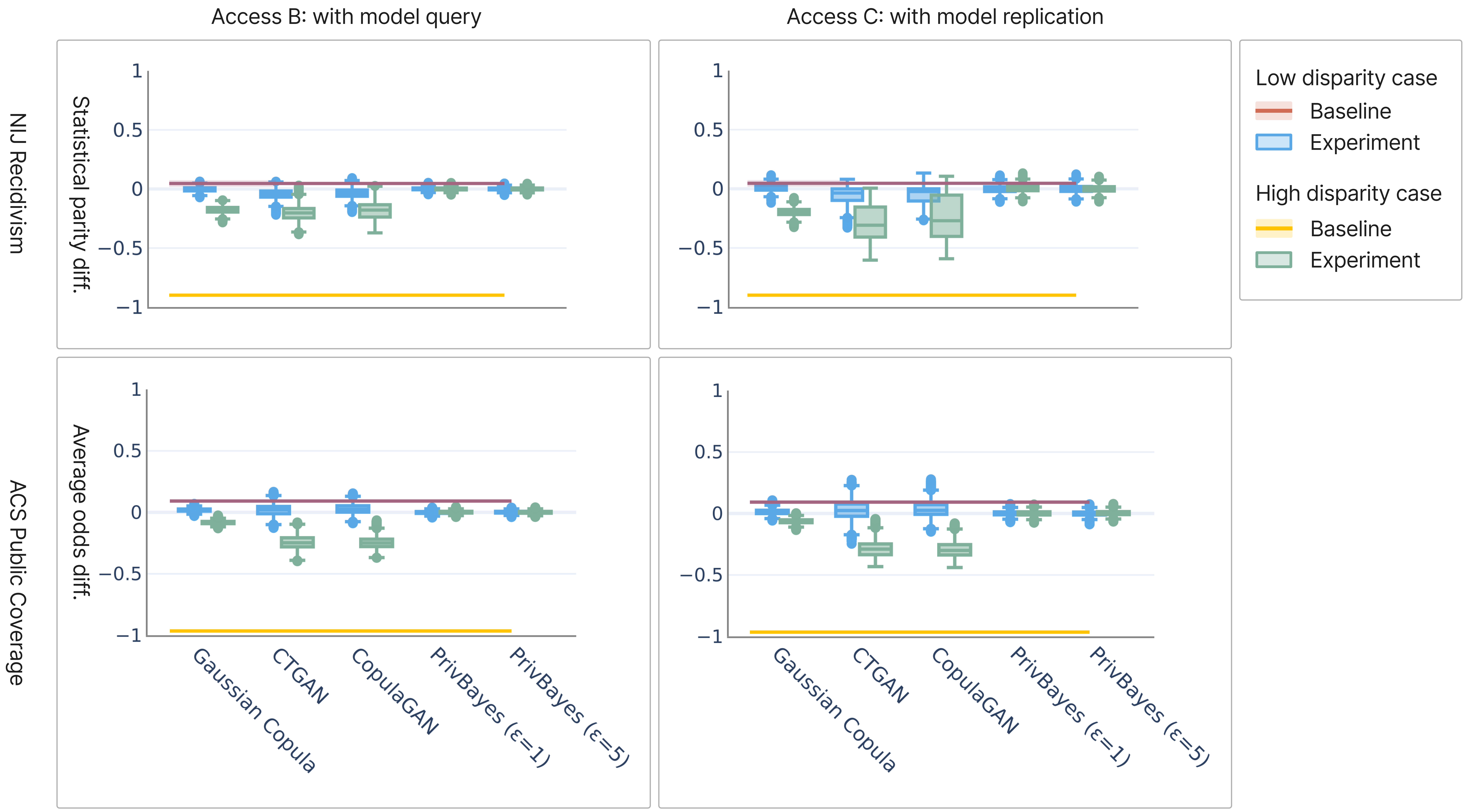}
\caption{\textbf{Metric reliability based on models used for synthetic data generation}, for Scenario B (left) and Scenario C (right).\\
The red and yellow lines represent median values for baselines, while the blue and green box plots represent values from audits on synthetic samples, with the middle lines representing medians.}
\label{fig:exp5}
\Description[Box plots for metric estimations based on the model used to generate synthetic data]{The figure includes 2 subplots for each dataset; one for Access Scenario B (with model predictions) and one for Access Scenario C (with model replication). The synthetic data generation models on the x-axis are ``Gaussian Copula'', ``CTGAN'', ``CopulaGAN'', ``PrivBayes with epsilon=1'', and ``PrivBayes with epsilon=5''. For high disparity cases, none of the audits on synthetic data yield parity values that overlap with the de facto true values.}
\end{figure*}

\subsubsection{Removal of features}
Figure~\ref{fig:exp2} shows that, within a high-dimensional model, only the important features significantly affect parity metrics estimation. Our results suggest that low importance features can be removed from the audit set without affecting auditor interpretation. However, a single important feature missing from the dataset when querying the model can mislead auditors. For high disparity cases, missing features skew metric estimates toward zero, leading to an underestimation of the disparity or a Type 2 error.

The impact of feature removal on audit reliability depends on the combination of dataset and model. For the NIJ Recidivism dataset, for which predictions rest on few strong predictors, metrics become unreliable after 5-7 low importance features are removed. For the ACS dataset, for which predictions rest on a higher proportion of predictors, metrics become unreliable after only 2 to 3 low importance features are removed (see~\Cref{fig:exp2}).
\balance

Under Access Scenario C (using model replication), Figure~\ref{fig:exp2} shows a larger error rate in metric estimation due to the reduced audit sample size. In high disparity cases, metric values are systematically under-estimated, but remain consistent as the number of missing features increases. In these cases, estimates deviate severely enough to cause an interpretation error (Type 2 error) when the main predictors are removed. In low disparity cases, metric estimations similarly deviate but appear to be more resilient to the removal of low importance features as compared to Access Scenario B (Figure~\ref{fig:exp2}).

\subsubsection{Disparate incompleteness}
Under Access Scenario B, Figure~\ref{fig:exp3} shows that in cases where baseline metrics indicate parity, audit metrics over-estimate disparities if values are disproportionately missing from the underprivileged group. Reversely, where baseline metrics indicate high disparity, increasing the rate of missing values lowers signals of disparity for both datasets, leading to a minor under-estimation of disparities. 

Under Access Scenario C, for the NIJ dataset, we find that metric estimations are more resilient to disparate incompleteness in low disparity cases, as well as in high disparity cases for high fractions of missing values.
For the ACS dataset, we observe a more important gap in metric estimations for the high disparity cases.

Overall, Table~\ref{tab:sumtable_exp3} shows that even 1\% of missing values among the underprivileged group can undermine audit reliability, with only respectively 72\% (Access Scenario B) and 25\% (Access Scenario C) of metric values within the baseline interval. While the interpretation of metrics can remain unaffected, our results suggest that disparity in missing data should be accounted for by auditors with individual-level data access.

\subsubsection{Synthetic data generation.}
We find that using synthetic data in place of real audit data significantly impacts audit reliability across metrics, datasets, disparity levels, and type of auditor access (\Cref{fig:exp5,tab:accessBC-sum}). Datasets generated by PrivBayes consistently fail to capture the disparities present in the original samples. Regardless of the true values, metrics computed on PrivBayes synthetic samples consistently suggest parity, with identical confidence intervals for estimates on low and high disparity cases. This pattern is evident in both the NIJ and ACS datasets. In contrast, the Gaussian Copula, CTGAN, and Copula GAN models do capture some differences between low and high disparity samples, as shown by the gaps between estimations on low-disparity and high-disparity cases (see \Cref{fig:exp5}). However, there is minimal overlap between the baseline and experimental intervals in low-disparity cases, and no overlap in high-disparity cases (see \Cref{tab:accessBC-sum}). In all cases where the baseline indicates group disparity (of any level), synthetic samples lead to an underestimation of that disparity, which may lead to Type 2 or ‘reverse’ errors.

\begin{table*}[!htbp]
  \centering
  \begin{tabular}{lllllcccccc}
    \toprule
    & & & & \multicolumn{6}{c}{Proportion of values within the baseline 95\% CI} \\ 
    
    Experiment & Dataset & Metric & Disparity & \multicolumn{3}{c}{Access B} & \multicolumn{3}{c}{Access C} \\ \midrule
    
    \addlinespace
    Subsampling & & & & \textit{n} = 3\% & \textit{n} = 20\% & \textit{n} = 30\% & \textit{n} = 3\% & \textit{n} = 20\% & \textit{n} = 30\% \\
    \cmidrule(l){5-7} \cmidrule(l){8-10}
    & \multirow{2}{*}{ACS} & \multirow{2}{*}{AOD} & High & {\color[HTML]{EA4335} 0.20} & 0.52 & 0.64 & {\color[HTML]{EA4335} 0.20} & 0.54 & {\color[HTML]{34A853} 0.70} \\
    & & & Low & {\color[HTML]{EA4335} 0.23} & 0.59 & {\color[HTML]{34A853} 0.70} & {\color[HTML]{EA4335} 0.25} & 0.56 & 0.65 \\
    \cmidrule[.1pt](l){2-10}
    & \multirow{2}{*}{NIJ} & \multirow{2}{*}{SPD} & High & {\color[HTML]{EA4335} 0.18} & 0.50 & 0.65 & {\color[HTML]{EA4335} 0.19} & 0.52 & 0.68 \\
    & & & Low & {\color[HTML]{EA4335} 0.24} & 0.59 & {\color[HTML]{34A853} 0.72} & {\color[HTML]{EA4335} 0.26} & 0.58 & {\color[HTML]{34A853} 0.70} \\ 
    \addlinespace
    \midrule
    \addlinespace

    Missing features & & & & \textit{f} = 12 & \textit{f} = 6 & \textit{f} = 1 & \textit{f} = 12 & \textit{f} = 6 & \textit{f} = 1 \\
    \cmidrule(l){5-7} \cmidrule(l){8-10}
    
    & \multirow{2}{*}{ACS} & \multirow{2}{*}{AOD} & High & {\color[HTML]{EA4335} <0.01} & {\color[HTML]{EA4335} <0.01} & 0.51 & {\color[HTML]{EA4335} <0.01} & {\color[HTML]{EA4335} <0.01} & {\color[HTML]{EA4335} <0.01} \\
    & & & Low & {\color[HTML]{EA4335} <0.01} & {\color[HTML]{EA4335} 0.05} & 0.61 & {\color[HTML]{EA4335} 0.03} & 0.31 & 0.32 \\
    \cmidrule[.1pt](l){2-10}
    & \multirow{2}{*}{NIJ} & \multirow{2}{*}{SPD} & High & {\color[HTML]{EA4335} <0.01} & {\color[HTML]{EA4335} <0.01} & {\color[HTML]{EA4335} 0.12} & {\color[HTML]{EA4335} <0.01} & {\color[HTML]{EA4335} <0.01} & {\color[HTML]{EA4335} <0.01} \\
    & & & Low & 0.54 & {\color[HTML]{34A853} 0.92} & {\color[HTML]{34A853} 0.93} & 0.63 & 0.65 & 0.66 \\ 
    \addlinespace
    \midrule
    \addlinespace

    Missing values & & & & \textit{m} = 20\% & \textit{m} = 5\% & \textit{m} = 1\% & \textit{m} = 20\% & \textit{m} = 5\% & \textit{m} = 1\% \\
    \cmidrule(l){5-7} \cmidrule(l){8-10}
    & \multirow{2}{*}{ACS} & \multirow{2}{*}{AOD} & High & {\color[HTML]{EA4335} <0.01} & {\color[HTML]{EA4335} 0.05} & 0.49 & {\color[HTML]{EA4335} <0.01} & {\color[HTML]{EA4335} <0.01} & {\color[HTML]{EA4335} <0.01} \\
    & & & Low & {\color[HTML]{EA4335} 0.24} & 0.50 & 0.61 & {\color[HTML]{EA4335} 0.11} & {\color[HTML]{EA4335} 0.30} & 0.33 \\
    \cmidrule[.1pt](l){2-10}
    & \multirow{2}{*}{NIJ} & \multirow{2}{*}{SPD} & High & {\color[HTML]{EA4335} <0.01} & {\color[HTML]{EA4335} <0.01} & {\color[HTML]{34A853} 0.86} & {\color[HTML]{EA4335} <0.01} & {\color[HTML]{EA4335} <0.01} & {\color[HTML]{EA4335} <0.01} \\
    & & & Low & {\color[HTML]{EA4335} 0.04} & {\color[HTML]{34A853} 0.73} & {\color[HTML]{34A853} 0.95} & 0.64 & {\color[HTML]{34A853} 0.71} & 0.65 \\ 
    \addlinespace
    \midrule
    \addlinespace

    Synthetic data & & & & CopGAN & CTGAN & PrBayes & CopGAN & CTGAN & PrBayes \\
    \cmidrule(l){5-7} \cmidrule(l){8-10}
    & \multirow{2}{*}{ACS} & \multirow{2}{*}{AOD} & High & {\color[HTML]{EA4335} <0.01} & {\color[HTML]{EA4335} <0.01} & {\color[HTML]{EA4335} <0.01} & {\color[HTML]{EA4335} <0.01} & {\color[HTML]{EA4335} <0.01} & {\color[HTML]{EA4335} <0.01} \\
    & & & Low & {\color[HTML]{EA4335} 0.07} & {\color[HTML]{EA4335} 0.06} & {\color[HTML]{EA4335} <0.01} & {\color[HTML]{EA4335} 0.08} & {\color[HTML]{EA4335} 0.07} & {\color[HTML]{EA4335} <0.01} \\
    \cmidrule[.1pt](l){2-10}
    & \multirow{2}{*}{NIJ} & \multirow{2}{*}{SPD} & High & {\color[HTML]{EA4335} <0.01} & {\color[HTML]{EA4335} <0.01} & {\color[HTML]{EA4335} <0.01} & {\color[HTML]{EA4335} <0.01} & {\color[HTML]{EA4335} <0.01} & {\color[HTML]{EA4335} <0.01} \\
    & & & Low & {\color[HTML]{EA4335} 0.06} & {\color[HTML]{EA4335} 0.03} & {\color[HTML]{EA4335} 0.07} & {\color[HTML]{EA4335} 0.06} & {\color[HTML]{EA4335} 0.03} & {\color[HTML]{EA4335} 0.07} \\ 

    \bottomrule
    \addlinespace
  \end{tabular}
  \caption{Proportion of values within the baseline 95\% confidence interval across experiments, datasets and access scenario. For the sub-sampling experiment, \textit{n} represents the shared sample size as compared with the baseline sample size. For the missing features experiment, \textit{f} indicates the number of features removed from the audit dataset. For the missing values experiment, \textit{m} indicates the proportion of values removed from the underprivileged group data (for the top 5 features only).}
  \label{tab:accessBC-sum}
\end{table*}

\subsection{Auditing Differentially Private models}
\label{results_dprf}
\begin{figure*}[!ht]
\centering
\includegraphics*[width=0.88\linewidth]{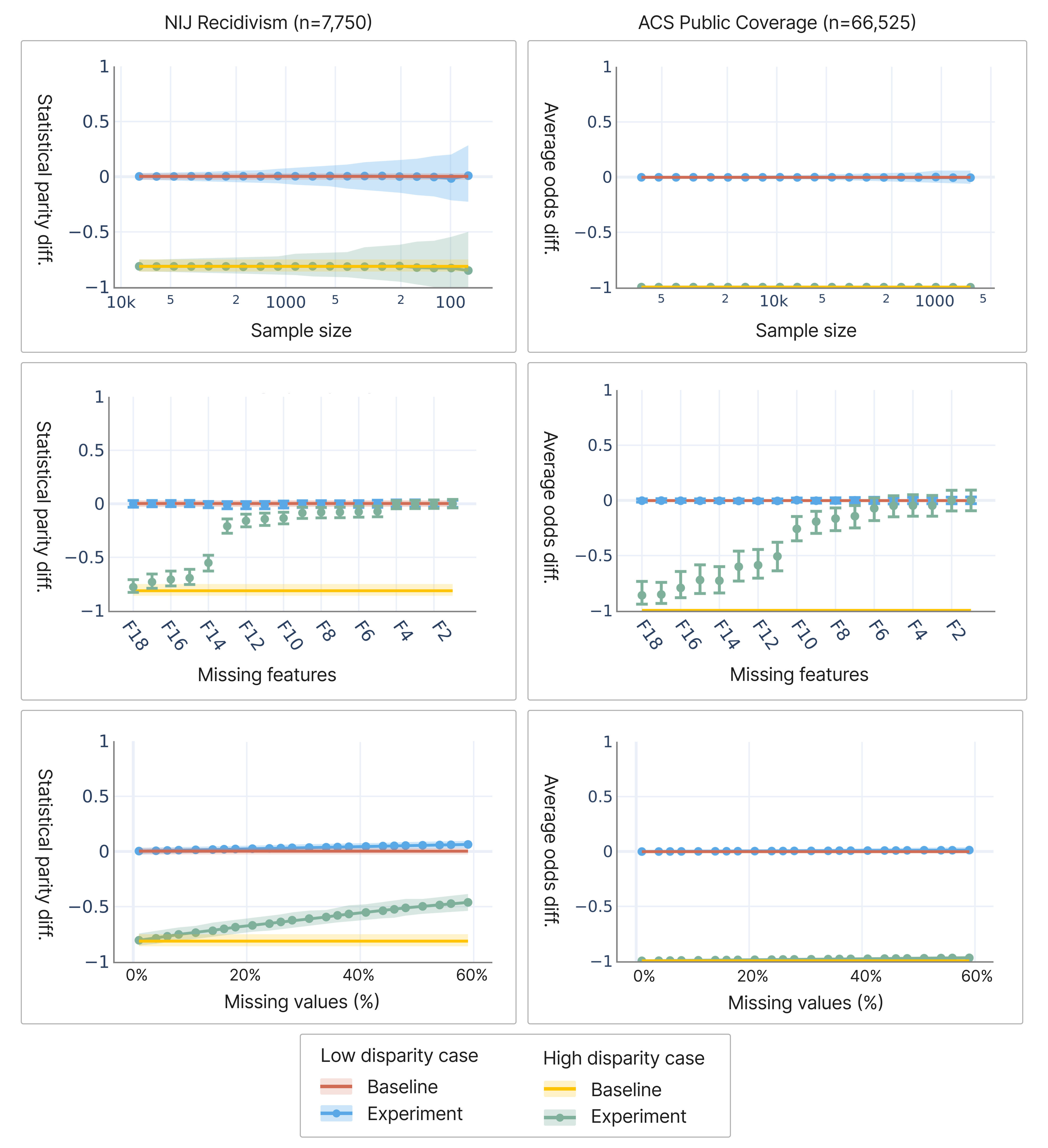}
\caption{Results for DP-RF model audits, for the NIJ dataset (left) and ACS dataset (right). The models were trained with $\epsilon$=10. The first row presents results for the reduced sample size experiment; the second row presents results for the missing features experiment; the third row presents results for the disparate missing values experiment.}
\label{fig:dprf}
\Description[Results for audits conducted on DP-RF models, for the NIJ dataset and the ACS dataset]{Subplots present results for models trained with $\epsilon$=10. The first row of subplots presents results for the reduced sample size experiment; the second row presents results for the missing features experiment; the third row presents results for the disparate missing values experiment. In all subplots, the y axis represents the metric value. For the NIJ dataset plots, baseline confidence intervals are visibly wider than in baseline audits on non-DP models. For the missing features experiment and high disparity cases, values deviate from the baseline earlier (i.e. at a lesser number of missing features) than when auditing non-DP models.}
\end{figure*}

\Cref{fig:dprf} shows the audit reliability under Scenario B for a Differentially Private Random Forest model set at $\epsilon=10$, for experiments (1) reduced sample size, (2) missing features, and (3) disparate missing values. At equal sample sizes, we find that baseline confidence intervals for parity metrics are wider on DP models than on non-DP models (more visible on the NIJ dataset), suggesting the need for larger audit samples. Sub-sampling and missing values experiments yield similar results across DP models with different privacy budgets and their non-DP counterparts. However, for both datasets, results diverge on  the missing features experiment: under high disparity conditions, DP models show greater deviations from baseline intervals (\Cref{fig:dprf}) compared to non-DP models (\Cref{fig:exp2}), irrespective of the privacy budget. This suggests that auditing DP models requires considering all predictors, regardless of feature importance, to ensure reliable metric estimates.

\subsection{Summary of findings and practical takeaways}

\subsubsection*{(A) When accessing aggregate statistics.}
\begin{itemize}
    \item Group parity metrics can be accurately computed from differentially private confusion matrices. The privacy budget ($\epsilon$ parameter) should not be set to very low values in cases where the sample is small, which can occur in highly imbalanced datasets and in intersectional fairness assessments. In general, the typically large sample sizes available to auditees allow some flexibility in setting the desired level of privacy protection, and privacy budgets can be conservative without compromising on audit reliability. 
    \item This type of access offers high metric accuracy for auditors, strong privacy protection for data subjects, and low disclosure from the auditee. Therefore, differentially-private confusion matrices are well-suited for public releases, potentially as a transparency requirement for organizations using decision-support algorithms, allowing initial diagnostics and some level of public accountability. However, these aggregates provide low flexibility for auditors' evaluations, and should not replace privileged access to individual-level data for more comprehensive audits. Further, releasing aggregates implies consulting with stakeholders, including independent researchers, to identify which statistics are of interest and what groups can be compared. Ideally, the availability of this data should be a default, rather than depending on Freedom of Information requests.\\
    Recently, Chen et al.~\cite{chen2024} used a ``remote data science'' tool---where auditees similarly returned differentially-private aggregations to auditors---to evaluate recommendation systems, demonstrating the value that can be derived from this type of access provided that auditees are cooperative and that auditors have sufficient flexibility in their queries.
    \item Auditors should be informed about the sample sizes and privacy budget used by the organization in order to evaluate the level of privacy protection and the reliability of metrics.
\end{itemize}

\newpage
\subsubsection*{(B) When accessing individual-level data with model predictions.} 
\begin{itemize}
    \item Our findings show that audits can become unreliable when samples are small ($n<1,000$), key predictive features are missing, or missing values are disproportionately found among some demographic groups. In our experiments, we examined these factors in isolation. If several of these issues were combined, datasets would likely become unusable for audits.
    \item Conversely, with reasonable sample size and complete data, it is highly unlikely that metric estimations would lead to interpretation errors. Although values may fall outside baseline confidence intervals, they generally remain within the same effect size range, preserving the accuracy of auditor interpretations. 
    \item Auditors can evaluate the suitability of a dataset for evaluation with regards to sample size and value missing-ness, but can difficultly account for missing features when the list of model predictors is unknown---as is often the case. In the case of missing features, auditors risk conducting misleading evaluations (typically leading to under-estimations of disparity) due to being unaware that their audit dataset is incomplete.
    \item The risk of interpretation errors can be further minimized by auditors by computing multiple estimations (e.g. using bootstrapping) rather than relying on a single value~\cite{besse2018,kallus2022}. This is especially important when auditing differentially private models, where estimates tend to vary more.
    \item Synthetic data generation has a strong tendency to invisibilize disparities, often leading to Type 2 errors. Therefore, synthetic data generation is highly misleading for auditing, and synthetic data should not replace real data in fairness evaluations.
\end{itemize}

\subsubsection*{(C) When accessing individual-level data without model predictions.}
\begin{itemize}
    \item Audits conducted on replications of the audited model are reliable, provided that the audit sample size is increased compared to Access B to allow for the replication of the model and a sufficiently large audit sample. Requirements for data completeness otherwise remain the same as in Access B.
    \item While in theory this method presents a viable workaround for situations where auditors cannot query the model, the feasibility of this method is limited by the accuracy of information about the model (model class and hyper-parameters), the availability of a (non-synthetic) dataset that matches the original training data in structure and quality, and the availability of necessary computational resources (in the case of large-scale AI~\cite{chen2021}). Under less ideal conditions, this audit method is highly likely to produce misleading results and be counter-productive for auditors. Therefore, providing access to high-quality data with model outputs or query access should be prioritized.
\end{itemize}

\clearpage
\section{Discussion} 

\subsection{Data protection, privacy-enhancing technologies, and algorithm auditing} \label{discussion_privacy}
Our findings suggest that privacy protection and reliable audit access are not incompatible. However, establishing best practices and incentivizing safe data sharing is a work in progress, and auditors urgently need solutions to facilitate access and increase trust in sharing practices. Designing these solutions will require collaboration from HCI, AI and technical privacy researchers, as well as regulators and audit practitioners.
Ultimately, the burden should not be on independent auditors to consider privacy protection and safe data access. Rather, these safety requirements must be integrated and facilitated by data holders and guided by regulation. 

In light of our findings, we hereby discuss how data-level privacy protection can be applied when sharing data with auditors. 

We find that data minimization, in the form of reduced sample size and number of features, can be used by data curators, such as the audited organization itself or a third party, without affecting audit results. This applies as long as the sample size remains reasonable and all important features are preserved, but comes with important caveats. First, it is common for deployed models to include a large number of predictors without regularization, despite the fact that these models can achieve similar performance with only the few strongest predictors~\cite{dressel2018}. Data minimization is an important principle of data protection regulation, and the fact that parity estimation is unaffected by the removal of most features in the audit dataset (for both Access Scenarios B and C) indicates that data minimization could potentially be applied upstream in the Machine Learning pipeline. Second, while removing features from a dataset shared with auditors does not always affect the accuracy of parity metrics, it could still mislead qualitative assessments, where the full list of predictors in a model is a crucial piece of information.

We show that differentially private aggregated statistics can offer strong privacy protection and accurate parity metrics (Access Scenario A). Despite the introduction of noise, we find that parity metrics are generally unlikely to be inaccurate to the point of causing interpretation errors for auditors.
Researchers have recently explored how bias introduced by Differential Privacy can be corrected ad hoc to ensure valid statistical inference while maintaining formal privacy guarantees~\cite{evans2023}. These new methods offer a promising avenue to establish both safe and reliable external access. However, individual-level data and the flexibility it affords auditors in their analysis~\cite{stadler2022} is still desirable in many cases. 
Granular data can lead to more trustworthy audits and more useful findings, as auditors can go beyond initial diagnostics and provide insights on underlying mechanisms of bias. Therefore, we emphasize that parity metrics derived from differentially private confusion matrices are generally reliable and provide a way forward for transparency and data availability. However, they are insufficient from an accountability perspective and should complement, rather than replace, higher levels of data access.

For higher levels of access, synthetic data is often presented as a promising alternative that allows access to individual-level data while maintaining strong privacy protection. Importantly, our findings echo others in the field in warning against downstream effects of synthetic data~\cite{stadler2022,vanbreugel2023}. We find that even state-of-the-art synthetic data generation models such as PrivBayes are not a viable option for quantitative audits, as they systematically invisibilize disparities. We warn against the adoption of synthetic data for algorithmic fairness evaluations, especially in the absence of appropriate documentation and validity guarantees. 

Finally, we found that differentially private Machine Learning models could be reliably audited, provided the audit sample is complete and sufficiently large, with a small accuracy loss for the auditee's model. Importantly, the use of such ML models do not necessarily provide any privacy protection for individuals in the audit dataset, but at least protect individuals in the original training dataset. DP models can offer a privacy-preserving alternative for organizations willing to release model access but no individual-level data, e.g. if auditors have access to public records with similar distributions.

\subsection{Implications and opportunities for regulators}
Policy reports increasingly recognize the value of algorithm audits and call for increased oversight for AI~\cite{tabassi2023,europeancommission2021}. On the other hand, companies and public institutions alike maintain high secrecy around prediction-based decision-making. In industry, companies actively lobby to limit access for auditors~\cite{casper2024}, claim intellectual property or “gaming the system” concerns as justification for opacity~\cite{heaven2020,algorithmwatch2020a}, and use legal retaliations that routinely put third-party auditors at risk~\cite{raji2022,kayser-bril2021,levine2021}. In the public sector, information about decision-making algorithms is rarely shared despite Freedom of Information laws~\cite{algorithmwatch2020,mcdonald2020}, making access to a dataset to conduct any empirical assessment extremely difficult. For instance, the Netherlands government refused to allow a technical audit on its welfare fraud detection algorithm SyRI~\cite{algorithmwatch2020a}. Similarly, the UK government discontinued the use of a biased algorithm in immigration services in 2020 after a judicial review~\cite{dark2020}, but never complied in sharing any further information about the algorithm~\cite{mcdonald2020}. Kuziemski and Misuraca argue there is a contradiction between the public sector’s motivation to increase its own efficiency, the secrecy surrounding that process, and the core objective of protecting citizens~\cite{kuziemski2020}, indicating that self-regulation is insufficient. Yet, the risks associated with transparency echo those in other industries, such as the financial sector, where audits are nevertheless standardized, and can be mitigated through privileged and secure access for auditors~\cite{casper2024}. 

From a legal standpoint, the feasibility of increased data access for algorithm audits is supported by existing regulations. In the EU, the Digital Markets Act (DMA) and Digital Services Act (DSA) emphasize the importance of data access for auditing purposes. Although focused on Very Large Online Platforms (VLOP) and Very Large Online Search Engines (VLOSE), they may provide a blueprint for extending external oversight to other sectors, including public sector decision-making systems. 

The DSA mandates data access to vetted researchers, including academic researchers and civic society organizations, and subjection to independent audits at least yearly~\cite{edelson2023}. Audited organizations are required to cooperate by providing auditors with all relevant data and to publish audit results. Privacy concerns are addressed at the stage of publicly releasing results rather than by restricting auditor access. This legal framework shifts the burden of proof on auditees to justify access restrictions, preventing denials of access based solely on commercial interests. This also balances power asymmetries between auditors and auditees~\cite{edelson2023}. The types of data that can be accessed under this mandate include performance metrics, training data, and model source code. Similarly, a consortium of digital regulators in the UK have argued for the need for academics to be part of the algorithm auditing landscape, with appropriate privacy safeguards, including ``a third-party sandbox in which algorithms can be shared by their owners and analysed in a privacy-preserving manner by appropriate external parties''~\cite{informationcommissionersoffice2022}.

However, these regulatory regimes still lack concrete guidelines for operationalizing algorithm audits, specifically related to compliance with data protection laws. Under the DMA, data holders are required to provide anonymized data “without substantially degrading the quality or usefulness of the data for the purpose of the DMA”~\cite{edelson2023}, a standard that is hardly realistic. Edelson, Graef, and Lancieri~\cite{edelson2023} report that a combination of legal and technical measures can ensure the necessary level of data access is granted in compliance with privacy law. The listed measures include “limiting the range and detail of the data to which access is provided” and relying on synthetic data. Yet, our findings suggest that privacy-enhancing techniques, such as data minimization or synthetic data, can mislead auditors and invisibilize disparities between demographic groups. Rather, auditors require granular and accurate data to produce reliable assessments. Moreover, synthetic data is not necessarily exempt from GDPR compliance, as it may still present re-identification risks~\cite{beduschi2024,gadotti2024a}. This raises the challenge of defining what qualifies as “anonymous enough”. Beduschi~\cite{beduschi2024} emphasizes the need for transparency regarding privacy-protection measures, such as clearly labelling synthetic data as such and providing contextual information about its generation to auditors.\\

Looking ahead, the AI and Data Acts promise to broaden regulatory oversight beyond the current focus on online platforms, extending it to the public sector. This shift is long overdue, as decision-making systems in the public sector carry a comparable or greater potential for harm, yet have received less scrutiny compared to large online platforms. Notably, the AI Act includes provisions to enhance transparency, enabling the general public to be informed when AI systems are in use and creating a database of high-risk AI systems~\cite{edelson2023}. Such measures could streamline the complex administrative and investigative procedures auditors currently face. However, it remains to be seen whether these upcoming regulations will translate into practical and actionable modes of access for auditors. More work is needed to ensure the applicability of AI regulations~\cite{chen2021}, meaningful auditor access, as well as legal protection for third-party auditors~\cite{chen2021,longpre2024}.

In any case, it is worth emphasizing that data protection laws do not generally prohibit researchers from accessing data. In fact, many provisions are designed to facilitate access when necessary. Specifically, the data minimization principle (GDPR, Art 5(1)(c)) states that data should be `limited to what is necessary,' but also that it must be `adequate' for the purposes of processing. Just as this principle does not prevent data processing for training AI~\cite{shanmugam2022}, it should also not prevent it for auditing purposes. The granularity and utility of data should not be degraded to the point of becoming insufficient for conducting a reliable algorithm audit.

\subsection{Implications and opportunities for HCI researchers}
The HCI community continuously bridges gaps between theoretical research and practice~\cite{lee2021a,deng2022}, and identifies needs for regulation, guidance, and tools~\cite{madaio2022,deng2022}. HCI research has been an important driver of algorithm auditing methods and frameworks, in shaping best practices~\cite{metaxa2021} and in documenting the struggles of auditors in practice~\cite{deng2023}. Researchers have recently developed tools to facilitate algorithm audits, such as for LLM assessments~\cite{arawjo2024}, fairness metric selection~\cite{ruf2022} and code reviews~\cite{kery2019}. For user-facing algorithms, user-driven audit frameworks have been proposed~\cite{lam2023,shen2021,devos2022} and are increasingly adopted by practitioners~\cite{deng2023}. Non-user-facing systems, which often similarly present high risks for harms, and data access issues in general, may present less obvious opportunities for traditional HCI research focused on end-users and interfaces; however, we contend that HCI research still plays an important role even with these non-user facing systems. These approaches are important to evaluate how audits benefit from increased access, and how this access can be made possible in practice. Independent researchers are important users of fairness toolkits, but face different challenges as compared to industry practitioners~\cite{deng2022,lee2021a,holstein2019,madaio2022,poland2022,ruf2022} in accessing and ensuring the reliability of data. Ojewale et al.~\cite{ojewale2024} report a low number of open-source tools built to support data access, as compared to evaluation tools (i.e. fairness toolkits), which are critical for external auditors. Some work is emerging in this space; for instance, Chen et al.~\cite{chen2024} study how data holders and auditors can collaborate through interactive privacy-enhancing tools. Therefore, we see an opportunity for HCI research to help realize the potential (and mitigate the risks) of third-party auditing for decision-support algorithms.

\subsection{Towards socio-technical algorithm audits}
Increased access is key for accurate socio-technical algorithm audits. Towards that goal, researchers have taken more interdisciplinary approaches and worked towards improving organization-auditor collaboration. Such collaboration allows for auditors to have more control or transparency on data sharing mechanisms. For instance, Seidelin et al.~\cite{seidelin2022} conducted a cooperative audit of the Danish government’s model assessing risk of long-term unemployment, and called the CSCW and HCI research community to work with institutions in developing approaches to audit public sector algorithms. Similarly, Obermeyer et al.~\cite{obermeyer2019} collaborated with a hospital to audit a model assigning patients to health programs, allowing them to both assess and correct racial discrimination. In both of these cases, collaboration allowed researchers to both diagnose issues in the model and correct the system accordingly. In our view, developing technical methods to bypass opacity is complementary with efforts to increase access and collaboration.

Further, computing parity metrics is only the first and least invasive step in an algorithm assessment. Parity metrics are valuable audit tools, and quantitative approaches as a whole contribute to progress on policy goals~\cite{mitchell2021}. However, they are not self-sufficient~\cite{lee2020,yurrita2022}. Comprehensive, socio-technical audits require deeper and more varied modes of access.

As argued by Corbett-Davies et al.~\cite{corbett-davies2017}, in order to ensure an algorithm is ``fair'', the algorithm itself needs to be assessed along with its decisions. Additionally, models are increasingly used to support decision-making and do not operate in a vacuum~\cite{chen2021}; therefore, holistic audits explore how the model is integrated within the broader socio-technical system, including how model predictions influence final decisions and actions taken. Beyond metric requirements, this involves access to contextual information about the model, its features, and the decision-making behind its development~\cite{casper2024,kommiyamothilal2024}. This may include datasheets~\cite{gebru2021} and model cards~\cite{mitchell2019} as a starting point. Mothilal, Guha and Ahmed~\cite{kommiyamothilal2024} proposed a Responsible ML framework that includes systematic documentation of decisions and assumptions along the development process. Providing access to this contextual documentation to auditors would make assessments both more efficient and impactful (similar to audit trails in other industries). Yet, Bhat et al.~\cite{bhat2023} report that despite being highly referenced, the adoption of documentation processes such as model cards has been low, which impedes accountability and highlights a discrepancy between theoretical “good practices” and practitioners’ workflows. In the public sector, this discrepancy materializes in government agencies promoting transparency while failing to document their own algorithm use. HCI research can support policy efforts by identifying needs and challenges for both those who produce and use this documentation~\cite{hind2020}. To ensure audit reliability, data holders should document decisions and measures taken to anonymize the data, allowing auditors to evaluate its suitability for their assessments. Requirements for documentation, as well as legal requirements for data integrity, are also necessary to establish trust and mitigate the risks of organizations ``faking fairness'' by voluntary manipulating their data before sharing it to auditors~\cite{fukuchi2020} or manipulating discriminatory prediction explanations to make them seem fair~\cite{aivodji2019}.

Socio-technical audits, supported by access to high-quality granular data, documentation and interdisciplinary collaboration, are meaningful drivers of change. However, the present challenges in obtaining sufficient access, particularly for non-user facing systems, highlight the need for continued research and practical solutions for auditors.

\subsection{Limitations and Future Work} \label{limitations}
Third-party auditors can encounter more restrictive conditions than those considered in our study, where auditors are assumed to have partial or full access to ground truth and sensitive features of interest (see \Cref{setting}). Our setting excludes applications where sensitive features are not collected, or cannot be collected. We also note that in cases where ground truth is either not readily available or not shared, auditors have a restricted choice of parity metrics to choose from. For instance, auditors from Lighthouse Reports selected Statistical Parity Difference due to access to ground-truth data being denied on grounds of privacy~\cite{braun2023}. Flexibility with regard to metric selection is beyond the scope of this study but should similarly be considered. 

In addition, we limited ourselves to binary classifiers and binary demographic groups, while auditors may audit other types of models (e.g. multi-classifiers) and assess intersectional disparities (i.e. among more than two subgroups).

We believe that auditing multi-classifiers would require more granularity and higher data quality in audit datasets, which presents a significant challenge for third-party audits. Future work could extend our approach to metrics that handle multi-classifiers, as well as other metrics encoding such as individual fairness and other data quality dimensions.

We also do not consider the potential for other kinds of privacy-preserving computation, or more complex data governance arrangements between auditors, auditees, and third parties such as regulators. For instance, secure multi-party computation could potentially be used to undertake algorithm audits and certify model fairness properties~\cite{kilbertus2018}. Furthermore, introducing trusted third parties (e.g. regulators or NGOs, as discussed by Veale and Binns~\cite{veale2017}) to act as data stewards through the audit process may lessen the trade-offs between privacy and audit reliability highlighted here.

\subsubsection*{Extending to intersectionality and other types of models}
Auditors increasingly consider intersectionality~\cite{cabrera2019,morina2020,ghosh2021,boxer2023} by assessing potential disparities between subgroups based on several demographic variables. The process of computing parity metrics would be similar as in this study. However, we suspect intersectional assessments, as well as assessments of risk-scorers and multi-classifiers (as opposed to binary classifiers), would require even more granular and higher quality data. In our case studies, we look at binary groups with relatively balanced representation (NIJ dataset) or large sample sizes (ACS dataset). Even under optimal data quality, auditors still need at least several hundred data points for each group; this required sample size grows significantly as quality lowers (e.g. incomplete data). In practice, it can be challenging for auditors to access datasets with sufficient subgroup representation for reliable metric estimates. Similarly, when dealing with differentially private aggregations, compared groups could be too narrow (due to high intersectionality and/or low representation), especially at lower sample sizes, which can strongly undermine reliability when setting low privacy budgets. This presents an important challenge, as intersectional groups capture the most vulnerable or marginalized individuals who potentially need the highest privacy protection.

\section{Conclusion}
This study assessed the reliability of algorithm audits under different conditions of data access. Our findings suggest that fairness audits with strong privacy guarantees are feasible under certain trade-offs, such as sacrificing granularity for differentially private aggregated statistics.

In addition to audit reliability and privacy protection, organizations and auditors must consider the flexibility provided by data access. Socio-technical assessments require higher levels of access with more granular data and auxiliary information. Yet, the level of access needed to ensure the reliability of even basic parity metrics---the foundation for more comprehensive audits---remains challenging for independent auditors to obtain. Current policies fall short of establishing robust oversight mechanisms for decision-making algorithms.
The HCI community has a unique opportunity to foster greater collaboration between auditors and organizations, enhancing data-sharing practices that benefit both fairness and privacy. Ultimately, improving data sharing practices can lead to more effective audits and, by extension, more equitable systems.



\bibliographystyle{acm}
\bibliography{references}

\appendix
\section{Appendices}
\label{sec:appendix}

    \subsection{Summary tables}
The tables below include results for three different group parity metrics: \textit{Average Odds Difference} (AOD), \textit{Statistical Parity Difference} (SPD), and \textit{Equal Opportunity Difference} (EOD). 

\begin{table*}[htbp]
  \centering
  \begin{tabular}{lll >{\centering\arraybackslash}m{1.2cm} >{\centering\arraybackslash}m{1.2cm} >{\centering\arraybackslash}m{1.2cm} | >{\centering\arraybackslash}m{1.2cm} >{\centering\arraybackslash}m{1.2cm} >{\centering\arraybackslash}m{1.2cm}}
    \toprule
    & & & \multicolumn{6}{c}{\begin{tabular}[c]{@{}c@{}}Proportion of values within the baseline CI at $x$\% subsampling\end{tabular}} \\ 
    & & & \multicolumn{3}{c}{(B) with model outputs} & \multicolumn{3}{c}{(C) with model replication} \\ \cmidrule(l){4-6} \cmidrule(l){7-9}
    \textbf{Case} & \textbf{Disparity level} & \textbf{Metric} & \textbf{3\%} & \textbf{30\%} & \textbf{80\%} & \textbf{3\%} & \textbf{20\%} & \textbf{30\%} \\ \midrule
    & & AOD & {\color[HTML]{EA4335}0.20} & 0.64 & {\color[HTML]{34A853}0.89} & {\color[HTML]{EA4335}0.20} & 0.54 & 0.70 \\
    & & SPD & {\color[HTML]{EA4335}0.19} & 0.59 & {\color[HTML]{34A853}0.88} & {\color[HTML]{EA4335}0.19} & 0.51 & 0.68 \\
    & \multirow{-3}{*}{High} & EOD & {\color[HTML]{EA4335}0.21} & 0.65 & {\color[HTML]{34A853}0.89} & {\color[HTML]{EA4335}0.22} & 0.55 & 0.69 \\
    \cmidrule(l){2-9}
    & & AOD & {\color[HTML]{EA4335}0.23} & 0.70 & {\color[HTML]{34A853}0.91} & {\color[HTML]{EA4335}0.25} & 0.56 & 0.65 \\
    & & SPD & {\color[HTML]{EA4335}0.23} & {\color[HTML]{34A853}0.71} & {\color[HTML]{34A853}0.91} & {\color[HTML]{EA4335}0.26} & 0.56 & 0.67 \\
    \multirow{-6.5}{*}{\begin{tabular}[c]{@{}l@{}}ACS Public\\Coverage\end{tabular}} & \multirow{-3}{*}{Low} & EOD & {\color[HTML]{EA4335}0.23} & 0.70 & {\color[HTML]{34A853}0.91} & {\color[HTML]{EA4335}0.24} & 0.58 & 0.69 \\
    \cmidrule(l){1-9}
    & & AOD & {\color[HTML]{EA4335}0.19} & 0.68 & {\color[HTML]{34A853}0.92} & {\color[HTML]{EA4335}0.20} & 0.53 & 0.69 \\
    & & SPD & {\color[HTML]{EA4335}0.18} & 0.65 & {\color[HTML]{34A853}0.90} & {\color[HTML]{EA4335}0.19} & 0.52 & 0.68 \\
    & \multirow{-3}{*}{High} & EOD & {\color[HTML]{EA4335}0.20} & 0.69 & {\color[HTML]{34A853}0.92} & {\color[HTML]{EA4335}0.21} & 0.56 & 0.70 \\
    \cmidrule(l){2-9}
    \multirow{-2.5}{*}{NIJ Recidivism} & \multirow{3}{*}{Low} & AOD & {\color[HTML]{EA4335}0.24} & {\color[HTML]{34A853}0.72} & {\color[HTML]{34A853}0.91} & {\color[HTML]{EA4335}0.25} & 0.56 & 0.69 \\
    & & SPD & {\color[HTML]{EA4335}0.24} & {\color[HTML]{34A853}0.72} & {\color[HTML]{34A853}0.91} & {\color[HTML]{EA4335}0.26} & 0.58 & 0.70 \\
    & & EOD & {\color[HTML]{EA4335}0.25} & 0.70 & {\color[HTML]{34A853}0.92} & {\color[HTML]{EA4335}0.25} & 0.59 & 0.69 \\
    \cmidrule(l){1-9}
    \multicolumn{2}{l}{} & \textbf{Means} & \textbf{0.22} & \textbf{0.68} & \textbf{0.90} & \textbf{0.23} & \textbf{0.55} & \textbf{0.69} \\
    \bottomrule
  \end{tabular}
  \caption{\textbf{Proportion of metric values within the baseline interval across subsampling levels}, for Access Scenarios (B) and (C)\\
  Subsampling percentages are in relation to the baseline audit sample size ($n=66,525$ for the ACS dataset, $n=7,751$ for the NIJ dataset). In Access scenario (C), given that 70\% of these samples are used for re-training, we only have values for up to 30\% of the baseline sample.}
  \label{tab:sumtable_exp1}
\end{table*}

\begin{table*}[htbp]
  \centering
  \begin{tabular}{lll >{\centering\arraybackslash}m{1.2cm} >{\centering\arraybackslash}m{1.2cm} >{\centering\arraybackslash}m{1.2cm} | >{\centering\arraybackslash}m{1.2cm} >{\centering\arraybackslash}m{1.2cm} >{\centering\arraybackslash}m{1.2cm}}
    \toprule
    & & & \multicolumn{6}{c}{\begin{tabular}[c]{@{}c@{}}Proportion of values within the baseline CI with $x$ missing features\end{tabular}} \\ 
    & & & \multicolumn{3}{c}{(B) with model outputs} & \multicolumn{3}{c}{(C) with model replication} \\ \cmidrule(l){4-6} \cmidrule(l){7-9}
    \textbf{Case} & \textbf{Disparity level} & \textbf{Metric} & \textbf{12} & \textbf{6} & \textbf{1} & \textbf{12} & \textbf{6} & \textbf{1} \\ \midrule
    & & AOD & {\color[HTML]{EA4335}0.00} & {\color[HTML]{EA4335}0.00} & 0.51 & {\color[HTML]{EA4335}0.00} & {\color[HTML]{EA4335}0.00} & {\color[HTML]{EA4335}0.00} \\
    & & SPD & {\color[HTML]{EA4335}0.00} & {\color[HTML]{EA4335}0.00} & 0.47 & {\color[HTML]{EA4335}0.00} & {\color[HTML]{EA4335}0.00} & {\color[HTML]{EA4335}0.00} \\
    & \multirow{-3}{*}{High} & EOD & {\color[HTML]{EA4335}0.00} & {\color[HTML]{EA4335}0.00} & 0.53 & {\color[HTML]{EA4335}0.00} & {\color[HTML]{EA4335}0.00} & {\color[HTML]{EA4335}0.00} \\
    \cmidrule(l){2-9}
    & & AOD & {\color[HTML]{EA4335}0.00} & {\color[HTML]{EA4335}0.05} & 0.61 & {\color[HTML]{EA4335}0.03} & 0.31 & 0.32 \\
    & & SPD & {\color[HTML]{EA4335}0.00} & {\color[HTML]{EA4335}0.03} & 0.60 & {\color[HTML]{EA4335}0.00} & {\color[HTML]{EA4335}0.26} & 0.30 \\
    \multirow{-6.5}{*}{\begin{tabular}[c]{@{}l@{}}ACS Public\\Coverage\end{tabular}} & \multirow{-3}{*}{Low} & EOD & {\color[HTML]{EA4335}0.00} & {\color[HTML]{EA4335}0.19} & 0.59 & {\color[HTML]{EA4335}0.14} & 0.31 & 0.33 \\
    \cmidrule(l){1-9}
    & & AOD & {\color[HTML]{EA4335}0.00} & {\color[HTML]{EA4335}0.00} & {\color[HTML]{EA4335}0.16} & {\color[HTML]{EA4335}0.00} & {\color[HTML]{EA4335}0.00} & {\color[HTML]{EA4335}0.00} \\
    & & SPD & {\color[HTML]{EA4335}0.00} & {\color[HTML]{EA4335}0.00} & {\color[HTML]{EA4335}0.12} & {\color[HTML]{EA4335}0.00} & {\color[HTML]{EA4335}0.00} & {\color[HTML]{EA4335}0.00} \\
    & \multirow{-3}{*}{High} & EOD & {\color[HTML]{EA4335}0.00} & {\color[HTML]{EA4335}0.08} & {\color[HTML]{34A853}0.85} & {\color[HTML]{EA4335}0.00} & {\color[HTML]{EA4335}0.00} & {\color[HTML]{EA4335}0.00} \\
    \cmidrule(l){2-9}
    \multirow{-2.5}{*}{NIJ Recidivism} & \multirow{3}{*}{Low} & AOD & 0.54 & {\color[HTML]{34A853}0.92} & {\color[HTML]{34A853}0.94} & 0.63 & 0.67 & 0.66 \\
    & & SPD & 0.54 & {\color[HTML]{34A853}0.92} & {\color[HTML]{34A853}0.93} & 0.63 & 0.65 & 0.66 \\
    & & EOD & 0.70 & {\color[HTML]{34A853}0.93} & {\color[HTML]{34A853}0.94} & 0.63 & 0.67 & {\color[HTML]{34A853}0.71} \\
    \cmidrule(l){1-9}
    \multicolumn{2}{l}{} & \textbf{Means} & \textbf{0.15} & \textbf{0.26} & \textbf{0.60} & \textbf{0.17} & \textbf{0.24} & \textbf{0.25} \\
    \bottomrule
  \end{tabular}
  \caption{\textbf{Proportion of metric values within the baseline interval based on the number of missing features} (Access Scenarios B and C). Both datasets have 18 features in total.}
  \label{tab:sumtable_exp2}
\end{table*}

\begin{table*}[htbp]
  \centering
  \begin{tabular}{lll >{\centering\arraybackslash}m{1.2cm} >{\centering\arraybackslash}m{1.2cm} >{\centering\arraybackslash}m{1.2cm} | >{\centering\arraybackslash}m{1.2cm} >{\centering\arraybackslash}m{1.2cm} >{\centering\arraybackslash}m{1.2cm}}
    \toprule
    & & & \multicolumn{6}{c}{\begin{tabular}[c]{@{}c@{}}Proportion of values within the baseline CI at $x$\% missing values\end{tabular}} \\ 
    & & & \multicolumn{3}{c}{(B) with model outputs} & \multicolumn{3}{c}{(C) with model replication} \\ \cmidrule(l){4-6} \cmidrule(l){7-9}
    \textbf{Case} & \textbf{Disparity level} & \textbf{Metric} & \textbf{20\%} & \textbf{5\%} & \textbf{1\%} & \textbf{20\%} & \textbf{5\%} & \textbf{1\%} \\ \midrule
    & & AOD & {\color[HTML]{EA4335}0.00} & {\color[HTML]{EA4335}0.05} & 0.49 & {\color[HTML]{EA4335}0.00} & {\color[HTML]{EA4335}0.00} & {\color[HTML]{EA4335}0.00} \\
    & & SPD & {\color[HTML]{EA4335}0.00} & {\color[HTML]{EA4335}0.00} & 0.37 & {\color[HTML]{EA4335}0.00} & {\color[HTML]{EA4335}0.00} & {\color[HTML]{EA4335}0.00} \\
    & \multirow{-3}{*}{High} & EOD & {\color[HTML]{EA4335}0.01} & {\color[HTML]{EA4335}0.21} & 0.53 & {\color[HTML]{EA4335}0.00} & {\color[HTML]{EA4335}0.00} & {\color[HTML]{EA4335}0.00} \\
    \cmidrule(l){2-9}
    & & AOD & {\color[HTML]{EA4335}0.24} & 0.50 & 0.61 & {\color[HTML]{EA4335}0.11} & 0.30 & 0.33 \\
    & & SPD & {\color[HTML]{EA4335}0.25} & 0.51 & 0.60 & {\color[HTML]{EA4335}0.18} & 0.35 & 0.33 \\
    \multirow{-6.5}{*}{\begin{tabular}[c]{@{}l@{}}ACS Public\\Coverage\end{tabular}} & \multirow{-3}{*}{Low} & EOD & {\color[HTML]{EA4335}0.14} & 0.48 & 0.59 & 0.30 & 0.36 & 0.37 \\
    \cmidrule(l){1-9}
    & & AOD & {\color[HTML]{EA4335}0.00} & {\color[HTML]{EA4335}0.03} & {\color[HTML]{34A853}0.89} & {\color[HTML]{EA4335}0.00} & {\color[HTML]{EA4335}0.00} & {\color[HTML]{EA4335}0.00} \\
    & & SPD & {\color[HTML]{EA4335}0.00} & {\color[HTML]{EA4335}0.00} & {\color[HTML]{34A853}0.86} & {\color[HTML]{EA4335}0.00} & {\color[HTML]{EA4335}0.00} & {\color[HTML]{EA4335}0.00} \\
    & \multirow{-3}{*}{High} & EOD & {\color[HTML]{EA4335}0.00} & 0.44 & {\color[HTML]{34A853}0.93} & {\color[HTML]{EA4335}0.00} & {\color[HTML]{EA4335}0.00} & {\color[HTML]{EA4335}0.00} \\
    \cmidrule(l){2-9}
    \multirow{-2.5}{*}{NIJ Recidivism} & \multirow{3}{*}{Low} & AOD & {\color[HTML]{EA4335}0.05} & {\color[HTML]{34A853}0.76} & {\color[HTML]{34A853}0.95} & 0.64 & {\color[HTML]{34A853}0.71} & 0.64 \\
    & & SPD & {\color[HTML]{EA4335}0.04} & {\color[HTML]{34A853}0.73} & {\color[HTML]{34A853}0.95} & 0.64 & {\color[HTML]{34A853}0.71} & 0.65 \\
    & & EOD & {\color[HTML]{EA4335}0.02} & 0.69 & {\color[HTML]{34A853}0.94} & 0.66 & 0.70 & 0.68 \\
    \cmidrule(l){1-9}
    \multicolumn{2}{l}{} & \textbf{Means} & \textbf{0.06} & \textbf{0.37} & \textbf{0.72} & \textbf{0.21} & \textbf{0.26} & \textbf{0.25} \\
    \bottomrule
  \end{tabular}
  \caption{\textbf{Proportion of metric values within the baseline interval based on rates of missing values among the underprivileged group} (Access scenarios B and C)}
  \label{tab:sumtable_exp3}
\end{table*}

\begin{table*}[htbp]
  \centering
  \begin{tabular}{lll m{1.2cm} m{1.2cm} m{1.2cm}}
    \toprule
    & & & \multicolumn{3}{c}{\begin{tabular}[c]{@{}c@{}}Proportion of values within the \\ baseline CI at $\epsilon = x$\end{tabular}} \\ \cmidrule(l){4-6}
    \textbf{Case} & \textbf{Disparity level} & \textbf{Metric} & \textbf{0.01} & \textbf{0.05} & \textbf{0.10} \\ \midrule
    & & AOD & {\color[HTML]{EA4335}0.29} & {\color[HTML]{34A853}0.73} & {\color[HTML]{34A853}0.91} \\
    & & SPD & {\color[HTML]{EA4335}0.07} & 0.62 & {\color[HTML]{34A853}0.82} \\
    & \multirow{-3}{*}{High} & EOD & 0.43 & {\color[HTML]{34A853}0.88} & {\color[HTML]{34A853}0.99} \\
    \cmidrule(l){2-6}
    & & AOD & 0.68 & {\color[HTML]{34A853}0.99} & {\color[HTML]{34A853}1.00} \\
    & & SPD & {\color[HTML]{34A853}0.75} & {\color[HTML]{34A853}0.97} & {\color[HTML]{34A853}0.99} \\
    \multirow{-6.5}{*}{\begin{tabular}[c]{@{}l@{}}ACS Public Coverage\\($n=66,525$)\end{tabular}} & \multirow{-3}{*}{Low} & EOD & {\color[HTML]{34A853}0.83} & {\color[HTML]{34A853}0.99} & {\color[HTML]{34A853}0.99} \\
    \cmidrule(l){1-6}
    \cmidrule(l){2-6}
    & & AOD & {\color[HTML]{EA4335}0.17} & 0.66 & {\color[HTML]{34A853}0.93} \\
    & & SPD & {\color[HTML]{EA4335}0.17} & 0.62 & {\color[HTML]{34A853}0.88} \\
    & \multirow{-3}{*}{High} & EOD & {\color[HTML]{EA4335}0.24} & {\color[HTML]{34A853}0.81} & {\color[HTML]{34A853}0.93} \\
    \cmidrule(l){2-6}
    \multirow{-2.5}{*}{\begin{tabular}[c]{@{}l@{}}NIJ Recidivism\\($n=7,750$)\end{tabular}} & \multirow{3}{*}{Low} & 
    AOD & 0.36 & {\color[HTML]{34A853}0.94} & {\color[HTML]{34A853}1.00} \\
    & & SPD & 0.40 & {\color[HTML]{34A853}0.94} & {\color[HTML]{34A853}1.00} \\
    & & EOD & 0.35 & {\color[HTML]{34A853}0.91} & {\color[HTML]{34A853}0.96} \\
    \cmidrule(l){1-6}
    \multicolumn{2}{l}{} & \textbf{Means} & \textbf{0.40} & \textbf{0.84} & \textbf{0.95} \\
    \cmidrule[0.6pt]{1-6}
    & & AOD & {\color[HTML]{EA4335}0.00} & {\color[HTML]{EA4335}0.04} & {\color[HTML]{EA4335}0.00} \\
    & & SPD & {\color[HTML]{EA4335}0.01} & {\color[HTML]{EA4335}0.01} & {\color[HTML]{EA4335}0.05} \\
    & \multirow{-3}{*}{High} & EOD & {\color[HTML]{EA4335}0.02} & {\color[HTML]{EA4335}0.01} & {\color[HTML]{EA4335}0.05} \\
    \cmidrule(l){2-6}
    & & AOD & {\color[HTML]{EA4335}0.02} & {\color[HTML]{EA4335}0.06} & {\color[HTML]{EA4335}0.13} \\
    & & SPD & {\color[HTML]{EA4335}0.05} & {\color[HTML]{EA4335}0.06} & {\color[HTML]{EA4335}0.07} \\
    \multirow{-6.5}{*}{\begin{tabular}[c]{@{}l@{}}ACS Public Coverage\\($n=1,000$)\end{tabular}} & \multirow{-3}{*}{Low} & EOD & {\color[HTML]{EA4335}0.00} & {\color[HTML]{EA4335}0.07} & {\color[HTML]{EA4335}0.14} \\
    \cmidrule(l){1-6}
    & & AOD & {\color[HTML]{EA4335}0.04} & {\color[HTML]{EA4335}0.13} & {\color[HTML]{EA4335}0.18} \\
    & & SPD & {\color[HTML]{EA4335}0.05} & {\color[HTML]{EA4335}0.12} & {\color[HTML]{EA4335}0.17} \\
    & \multirow{-3}{*}{High} & EOD & {\color[HTML]{EA4335}0.08} & {\color[HTML]{EA4335}0.15} & {\color[HTML]{EA4335}0.28} \\
    \cmidrule(l){2-6}
    \multirow{-2.5}{*}{\begin{tabular}[c]{@{}l@{}}NIJ Recidivism\\
    ($n=1,000$)\end{tabular}} & \multirow{3}{*}{Low} & 
    AOD & {\color[HTML]{EA4335}0.06} & {\color[HTML]{EA4335}0.20} & 0.43 \\
    & & SPD & {\color[HTML]{EA4335}0.06} & {\color[HTML]{EA4335}0.28} & 0.39 \\
    & & EOD & {\color[HTML]{EA4335}0.13} & {\color[HTML]{EA4335}0.22} & 0.32 \\
    \cmidrule(l){1-6}
    \multicolumn{2}{l}{} & \textbf{Means} & \textbf{0.04} & \textbf{0.12} & \textbf{0.18} \\
    \bottomrule
  \end{tabular}
  \caption{\textbf{Proportion of metric values within the baseline interval based on $\epsilon$ parameter} (Access scenario A).\\Values above 0.70 are indicated in green, and values below 0.30 are indicated in red.}
  \label{tab:scenarioA_exp4}
\end{table*}

\begin{table*}[htbp]
  \centering
  \begin{tabular}{m{1.6cm} m{0.9cm} l m{1.2cm} m{1.2cm} m{1.2cm} m{1.2cm} m{1.2cm}}
    \toprule
    & & & \multicolumn{5}{c}{\begin{tabular}[c]{@{}c@{}}Proportion of values within the baseline CI\end{tabular}} \\ \cmidrule(l){4-8}
    \textbf{Case} & \textbf{Disparity level} & \textbf{Metric} & \textbf{Gaussian Copula} & \textbf{Copula GAN} & \textbf{CTGAN} & \textbf{PrivBayes ($\epsilon$=1)} & \textbf{PrivBayes ($\epsilon$=5)} \\ \midrule
    & & AOD & {\color[HTML]{EA4335}0.00} & {\color[HTML]{EA4335}0.00} & {\color[HTML]{EA4335}0.00} & {\color[HTML]{EA4335}0.00} & {\color[HTML]{EA4335}0.00} \\
    & & SPD & {\color[HTML]{EA4335}0.00} & {\color[HTML]{EA4335}0.00} & {\color[HTML]{EA4335}0.00} & {\color[HTML]{EA4335}0.00} & {\color[HTML]{EA4335}0.00} \\
    & & EOD & {\color[HTML]{EA4335}0.00} & {\color[HTML]{EA4335}0.00} & {\color[HTML]{EA4335}0.00} & {\color[HTML]{EA4335}0.00} & {\color[HTML]{EA4335}0.00} \\ 
    \cmidrule(l){2-8}
    & \multirow{-6}{*}{High} & AOD & {\color[HTML]{EA4335}0.00} & {\color[HTML]{EA4335}0.07} & {\color[HTML]{EA4335}0.06} & {\color[HTML]{EA4335}0.00} & {\color[HTML]{EA4335}0.00} \\
    & & SPD & {\color[HTML]{EA4335}0.00} & {\color[HTML]{EA4335}0.01} & {\color[HTML]{EA4335}0.01} & {\color[HTML]{EA4335}0.00} & {\color[HTML]{EA4335}0.00} \\ 
    \multirow{-6.5}{*}{\parbox[t]{1.7cm}{ACS Public\\Coverage}} & \multirow{-3}{*}{Low} & EOD & {\color[HTML]{EA4335}0.01} & {\color[HTML]{EA4335}0.15} & {\color[HTML]{EA4335}0.10} & {\color[HTML]{EA4335}0.00} & {\color[HTML]{EA4335}0.00} \\
    \cmidrule(l){1-8}
    & & AOD & {\color[HTML]{EA4335}0.00} & {\color[HTML]{EA4335}0.00} & {\color[HTML]{EA4335}0.00} & {\color[HTML]{EA4335}0.00} & {\color[HTML]{EA4335}0.00} \\
    & & SPD & {\color[HTML]{EA4335}0.00} & {\color[HTML]{EA4335}0.00} & {\color[HTML]{EA4335}0.00} & {\color[HTML]{EA4335}0.00} & {\color[HTML]{EA4335}0.00} \\
    & \multirow{-3}{*}{High} & EOD & {\color[HTML]{EA4335}0.00} & {\color[HTML]{EA4335}0.00} & {\color[HTML]{EA4335}0.00} & {\color[HTML]{EA4335}0.00} & {\color[HTML]{EA4335}0.00} \\
    \cmidrule(l){2-8}
    \multirow{-2.5}{*}{\parbox[t]{1.7cm}{NIJ\\Recidivism}} & \multirow{3}{*}{Low} & AOD & {\color[HTML]{EA4335}0.18} & {\color[HTML]{EA4335}0.08} & {\color[HTML]{EA4335}0.05} & {\color[HTML]{EA4335}0.13} & {\color[HTML]{EA4335}0.13} \\
    & & SPD & {\color[HTML]{EA4335}0.14} & {\color[HTML]{EA4335}0.06} & {\color[HTML]{EA4335}0.03} & {\color[HTML]{EA4335}0.07} & {\color[HTML]{EA4335}0.07} \\
    & & EOD & {\color[HTML]{34A853}0.73} & 0.34 & {\color[HTML]{EA4335}0.25} & {\color[HTML]{34A853}0.84} & {\color[HTML]{34A853}0.81} \\
    \cmidrule(l){1-8}
    \multicolumn{2}{l}{} & \textbf{Means} & \textbf{0.09} & \textbf{0.06} & \textbf{0.04} & \textbf{0.09} & \textbf{0.08} \\
    \bottomrule
  \end{tabular}
  \caption{\textbf{Proportion of metric values within the baseline interval based on synthetic data generation model}\\(Access scenario B)}
  \label{tab:scenarioB_exp5}
\end{table*}

\begin{table*}[htbp]
  \centering
  \begin{tabular}{m{1.6cm} m{0.9cm} l m{1.2cm} m{1.2cm} m{1.2cm} m{1.2cm} m{1.2cm}}
    \toprule
    & & & \multicolumn{5}{c}{\begin{tabular}[c]{@{}c@{}}Proportion of values within the baseline CI\end{tabular}} \\ \cmidrule(l){4-8}
    \textbf{Case} & \textbf{Disparity level} & \textbf{Metric} & \textbf{Gaussian Copula} & \textbf{Copula GAN} & \textbf{CTGAN} & \textbf{PrivBayes ($\epsilon$=1)} & \textbf{PrivBayes ($\epsilon$=5)} \\ \midrule
    & & AOD & {\color[HTML]{EA4335}0.00} & {\color[HTML]{EA4335}0.00} & {\color[HTML]{EA4335}0.00} & {\color[HTML]{EA4335}0.00} & {\color[HTML]{EA4335}0.00} \\
    & & SPD & {\color[HTML]{EA4335}0.00} & {\color[HTML]{EA4335}0.00} & {\color[HTML]{EA4335}0.00} & {\color[HTML]{EA4335}0.00} & {\color[HTML]{EA4335}0.00} \\
    & & EOD & {\color[HTML]{EA4335}0.00} & {\color[HTML]{EA4335}0.00} & {\color[HTML]{EA4335}0.00} & {\color[HTML]{EA4335}0.00} & {\color[HTML]{EA4335}0.00} \\ 
    \cmidrule(l){2-8}
    & \multirow{-6}{*}{High} & AOD & {\color[HTML]{EA4335}0.00} & {\color[HTML]{EA4335}0.08} & {\color[HTML]{EA4335}0.07} & {\color[HTML]{EA4335}0.00} & {\color[HTML]{EA4335}0.00} \\
    & & SPD & {\color[HTML]{EA4335}0.00} & {\color[HTML]{EA4335}0.04} & {\color[HTML]{EA4335}0.04} & {\color[HTML]{EA4335}0.00} & {\color[HTML]{EA4335}0.00} \\ 
    \multirow{-6.5}{*}{\parbox[t]{1.7cm}{ACS Public\\Coverage}} & \multirow{-3}{*}{Low} & EOD & {\color[HTML]{EA4335}0.08} & {\color[HTML]{EA4335}0.15} & {\color[HTML]{EA4335}0.12} & {\color[HTML]{EA4335}0.01} & {\color[HTML]{EA4335}0.01} \\
    \cmidrule(l){1-8}
    & & AOD & {\color[HTML]{EA4335}0.00} & {\color[HTML]{EA4335}0.00} & {\color[HTML]{EA4335}0.00} & {\color[HTML]{EA4335}0.00} & {\color[HTML]{EA4335}0.00} \\
    & & SPD & {\color[HTML]{EA4335}0.00} & {\color[HTML]{EA4335}0.00} & {\color[HTML]{EA4335}0.00} & {\color[HTML]{EA4335}0.00} & {\color[HTML]{EA4335}0.00} \\
    & \multirow{-3}{*}{High} & EOD & {\color[HTML]{EA4335}0.00} & {\color[HTML]{EA4335}0.00} & {\color[HTML]{EA4335}0.00} & {\color[HTML]{EA4335}0.00} & {\color[HTML]{EA4335}0.00} \\
    \cmidrule(l){2-8}
    \multirow{-2.5}{*}{\parbox[t]{1.7cm}{NIJ\\Recidivism}} & \multirow{3}{*}{Low} & AOD & 0.40 & {\color[HTML]{EA4335}0.04} & {\color[HTML]{EA4335}0.04} & {\color[HTML]{EA4335}0.29} & 0.32 \\
    & & SPD & 0.34 & {\color[HTML]{EA4335}0.03} & {\color[HTML]{EA4335}0.02} & {\color[HTML]{EA4335}0.24} & {\color[HTML]{EA4335}0.27} \\
    & & EOD & 0.63 & 0.42 & 0.40 & 0.48 & 0.46 \\
    \cmidrule(l){1-8}
    \multicolumn{2}{l}{} & \textbf{Means} & \textbf{0.12} & \textbf{0.06} & \textbf{0.06} & \textbf{0.09} & \textbf{0.09} \\
    \bottomrule
  \end{tabular}
  \caption{\textbf{Proportion of metric values within the baseline interval based on synthetic data generation model}\\(Access scenario C)}
  \label{tab:scenarioC_exp5}
\end{table*}

\clearpage

    \subsection{Experiments on the ACS Public Coverage dataset across three metrics}

\begin{figure*}[!htbp]
\centering
\includegraphics*[width=0.9\linewidth]{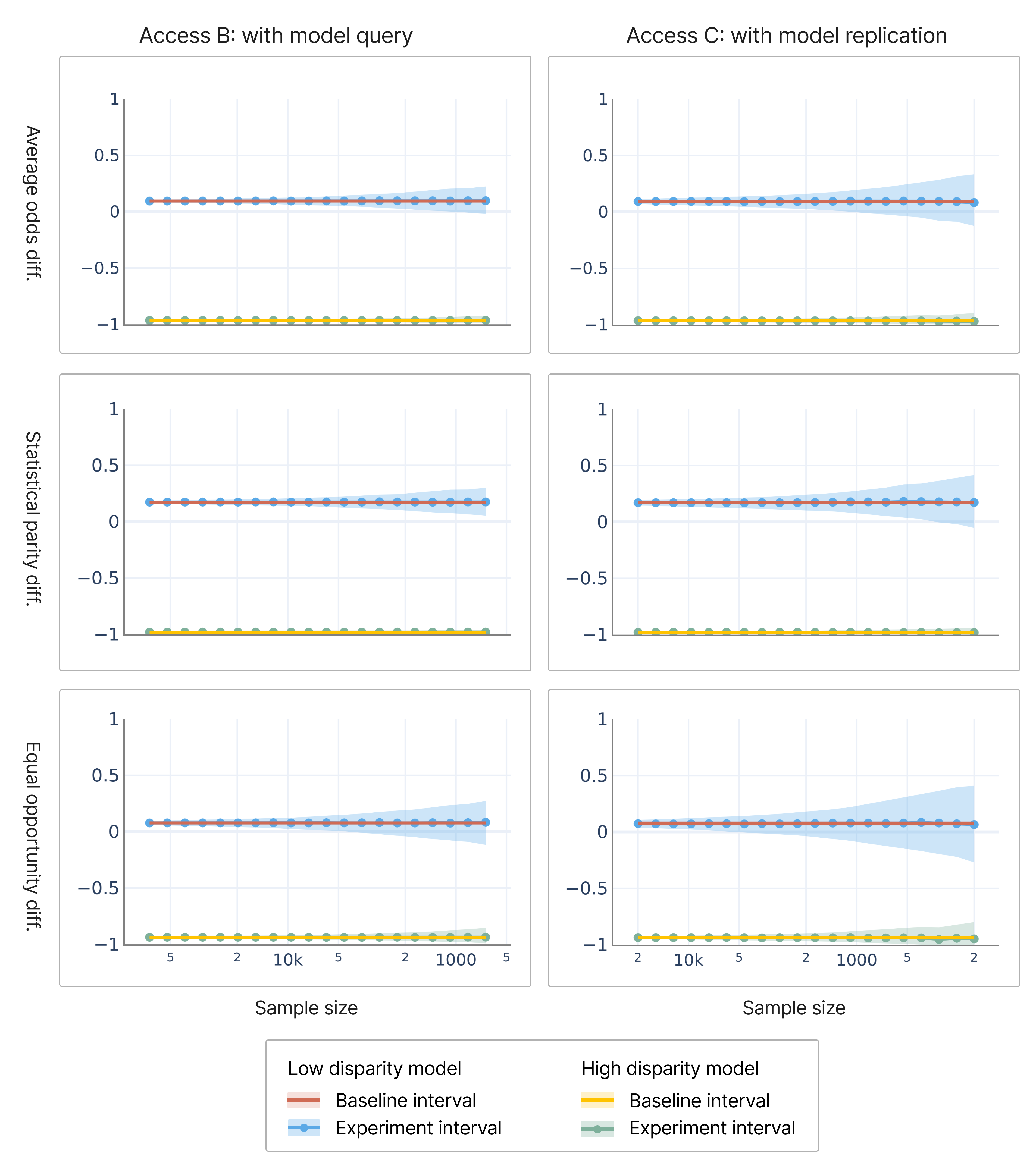}
\caption{\textbf{Effect of sample size on metric reliability} for Access B (left) and Access C (right) (ACS dataset)}
\label{fig:acs_exp1}
\Description[Line plots for metric estimations across audit sample size for the ACS Public Coverage dataset]{The figure includes 2 subplots for each parity metric; one for Access Scenario B (with model predictions) and one for Access Scenario C (with model replication). Each plot displays baseline confidence intervals and experiment intervals across sample sizes, for both low and high disparity cases. In all cases, the experiment intervals become broader as the sample size decreases. At equal sample sizes, the confidence interval ranges are similar between access B and C.}
\end{figure*}

\begin{figure*}[!htbp]
\centering
\includegraphics*[width=0.9\linewidth]{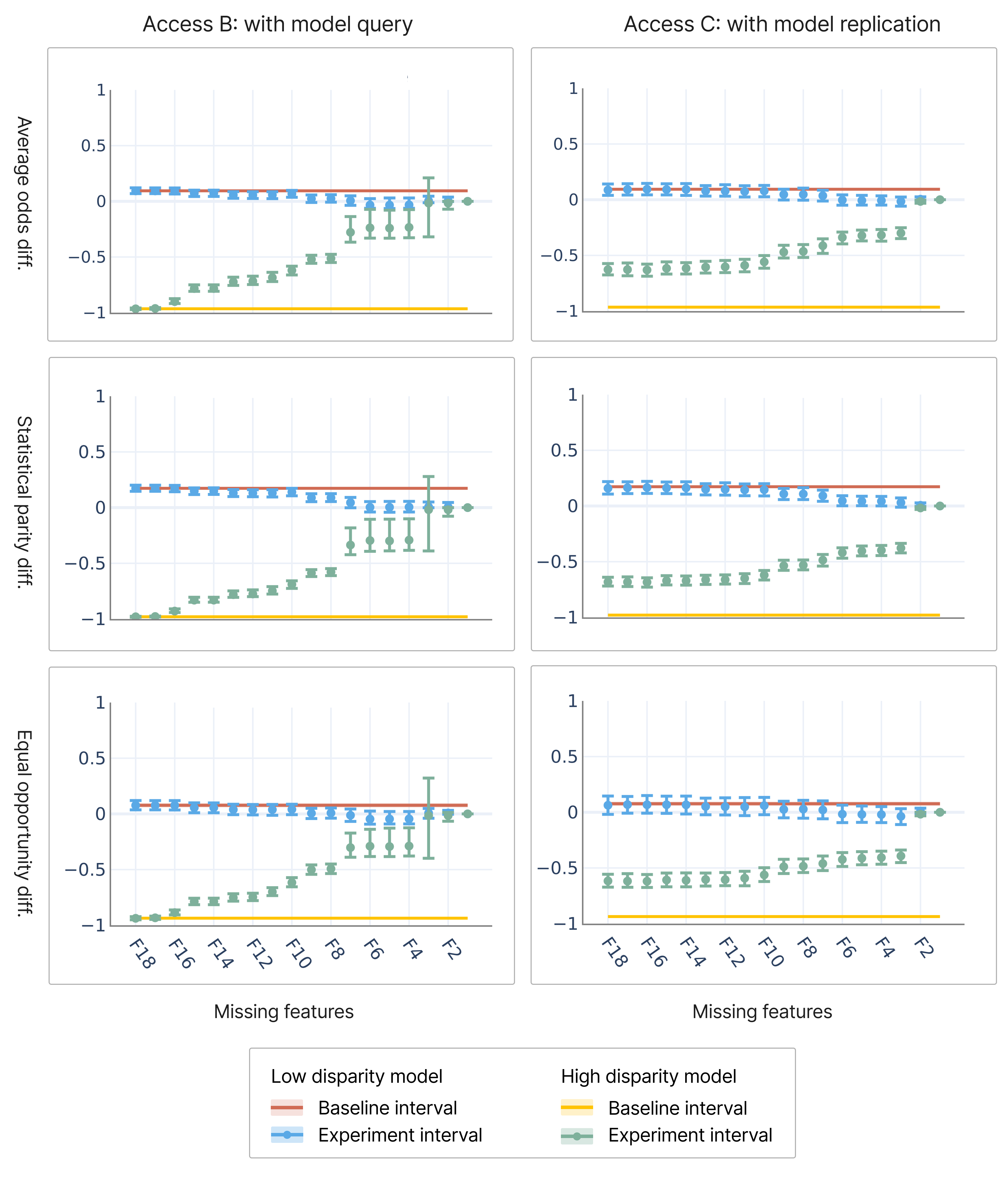}
\caption{\textbf{Effect of missing features on metric reliability} for Access B (left) and Access C (right) (ACS dataset)\\
On the x-axis, features are ordered by increasing order of importance. Plots are cumulative (e.g. at the F14 point, features 14 to 18 are missing from the audit dataset).}
\label{fig:acs_exp2}
\Description[Scatter plots for metric estimations based on feature removal, for the ACS Public Coverage dataset]{The figure includes 2 subplots for each parity metric; one for Access Scenario B (with model predictions) and one for Access Scenario C (with model replication). Each plot displays baseline confidence intervals and experiment intervals across sample sizes, for both low and high disparity cases. In all cases, the experiment values deviate towards 0 as the cumulative number of missing features increases.}
\end{figure*}

\begin{figure*}[!htbp]
\centering
\includegraphics*[width=0.9\linewidth]{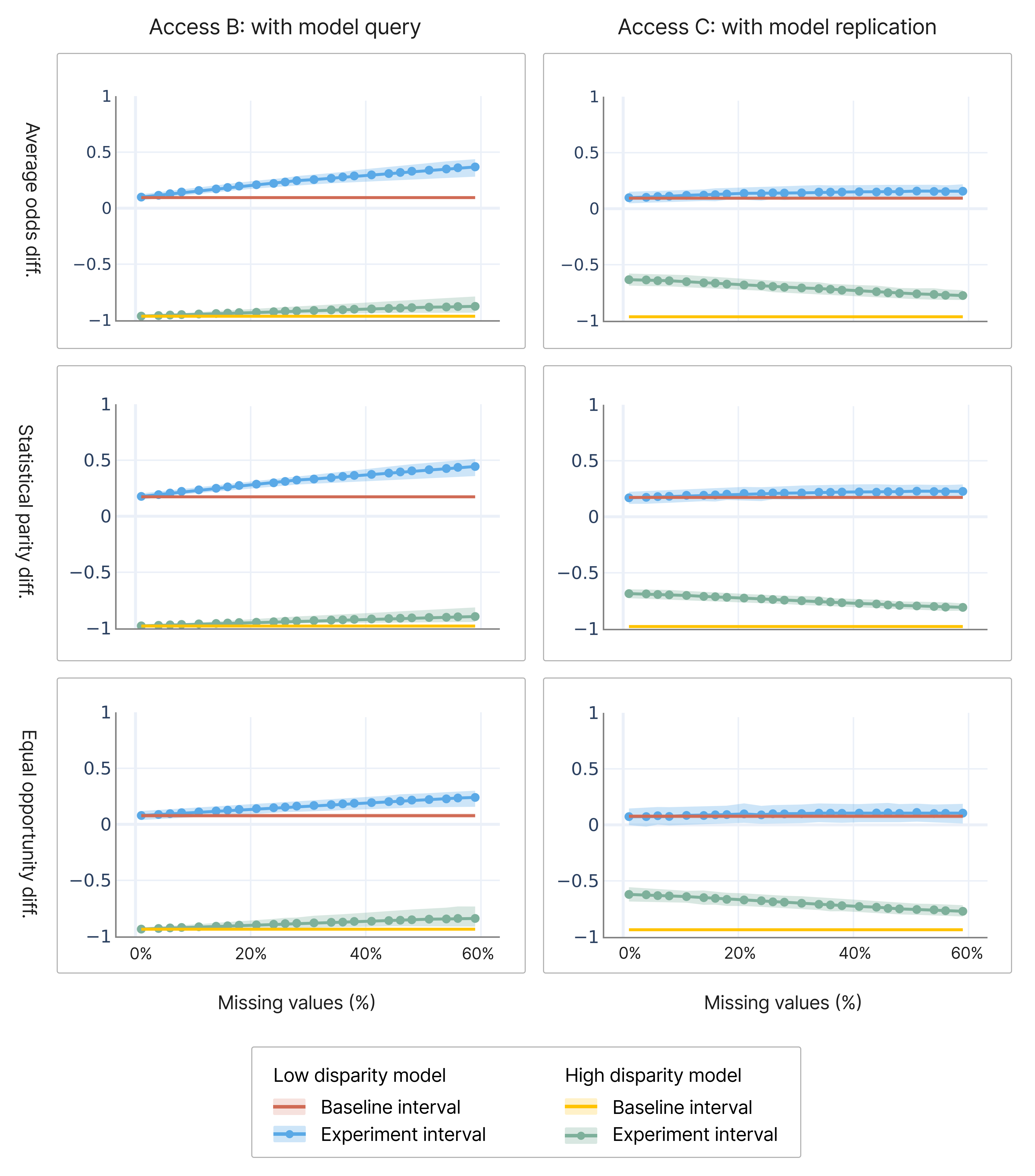}
\caption{\textbf{Effect of disparate missing values rates on metric reliability} for Access B (left) and Access C (right) (ACS dataset)}
\label{fig:acs_exp3}
\Description[Scatter plots for metric estimations based on rates of missing values among the underprivileged group, for the ACS Public Coverage dataset]{The figure includes 2 subplots for each parity metric; one for Access Scenario B (with model predictions) and one for Access Scenario C (with model replication). Each plot displays baseline confidence intervals and experiment intervals across sample sizes, for both low and high disparity cases. For access scenario B, experiment intervals deviate from the baseline as the rate of missing values increases. For access scenario C with low disparity, experiment intervals remain close to the baseline. For access scenario C with high disparity, experiment intervals start with an important underestimation of the disparity, and deviate toward the de facto true value as the missing rate increases.}
\end{figure*}

\begin{figure*}[!htbp]
\centering
\includegraphics*[width=0.9\linewidth]{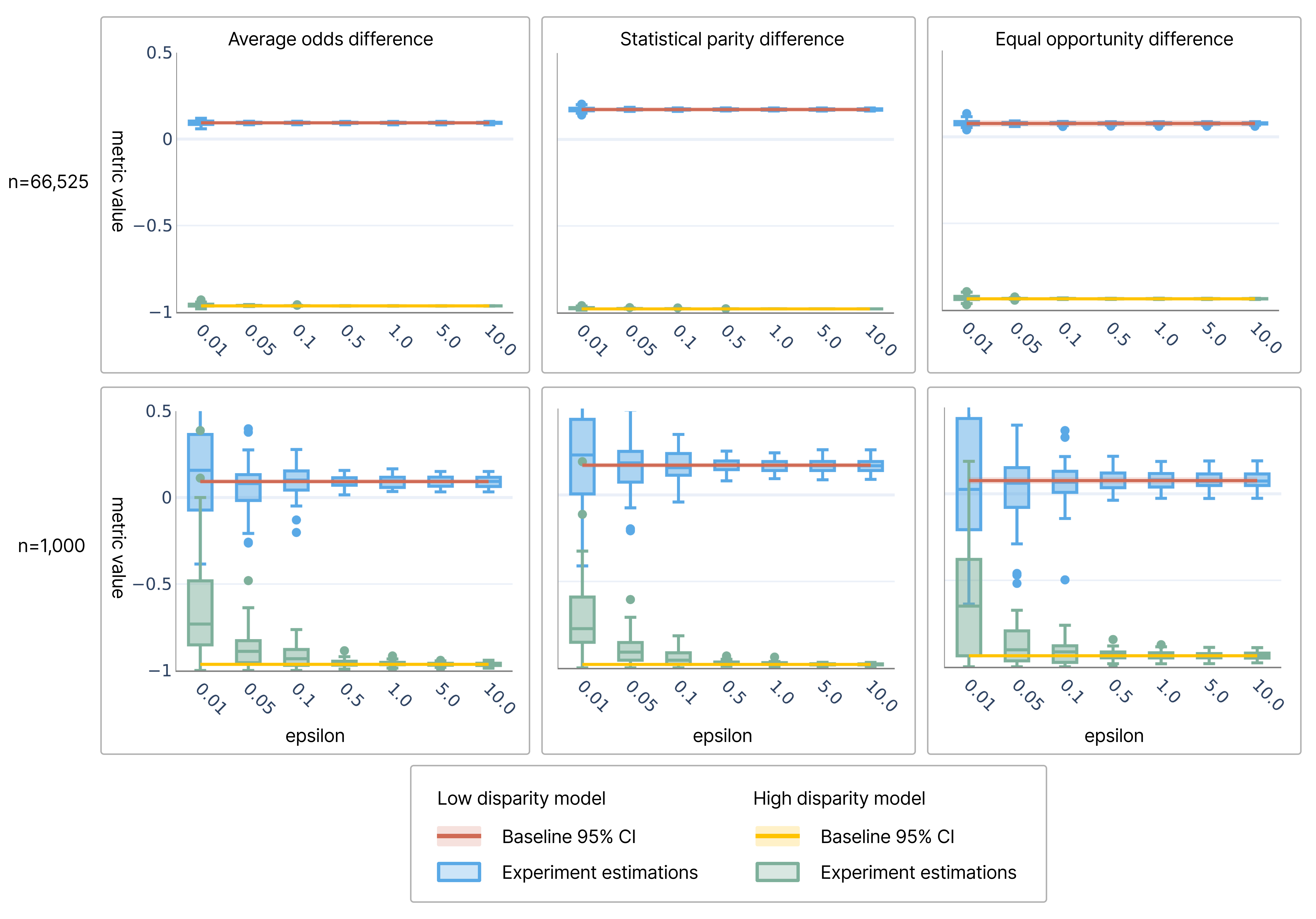}
\caption{\textbf{Effect of differential privacy on metric reliability} for the ACS dataset, at full sample size ($n=66,525$) and with a reduced sample size ($n=1,000$)}
\label{fig:acs_exp4}
\Description[Box plots for metric estimations across epsilon values for the ACS dataset, across two sample sizes (full sample size on the left, and n=1000 on the right]{The plots represent (a) confidence intervals for baseline audits and (b) box plots for experiment values, across a range of epsilon values (from 0.01 to 10), for both low and high disparity cases, and for three group parity metrics: Average Odds Difference, Statistical Parity Difference, and Equal Opportunity Difference. At full sample sizes, box plots have narrow intervals and high overlap with the baselines. For n=1000, box plots are very large at epsilon=0.01, and their size gradually reduces as epsilon is increased. These patterns replicate across all three parity metrics.}
\end{figure*}

\begin{figure*}[!htbp]
\centering
\includegraphics*[width=0.9\linewidth]{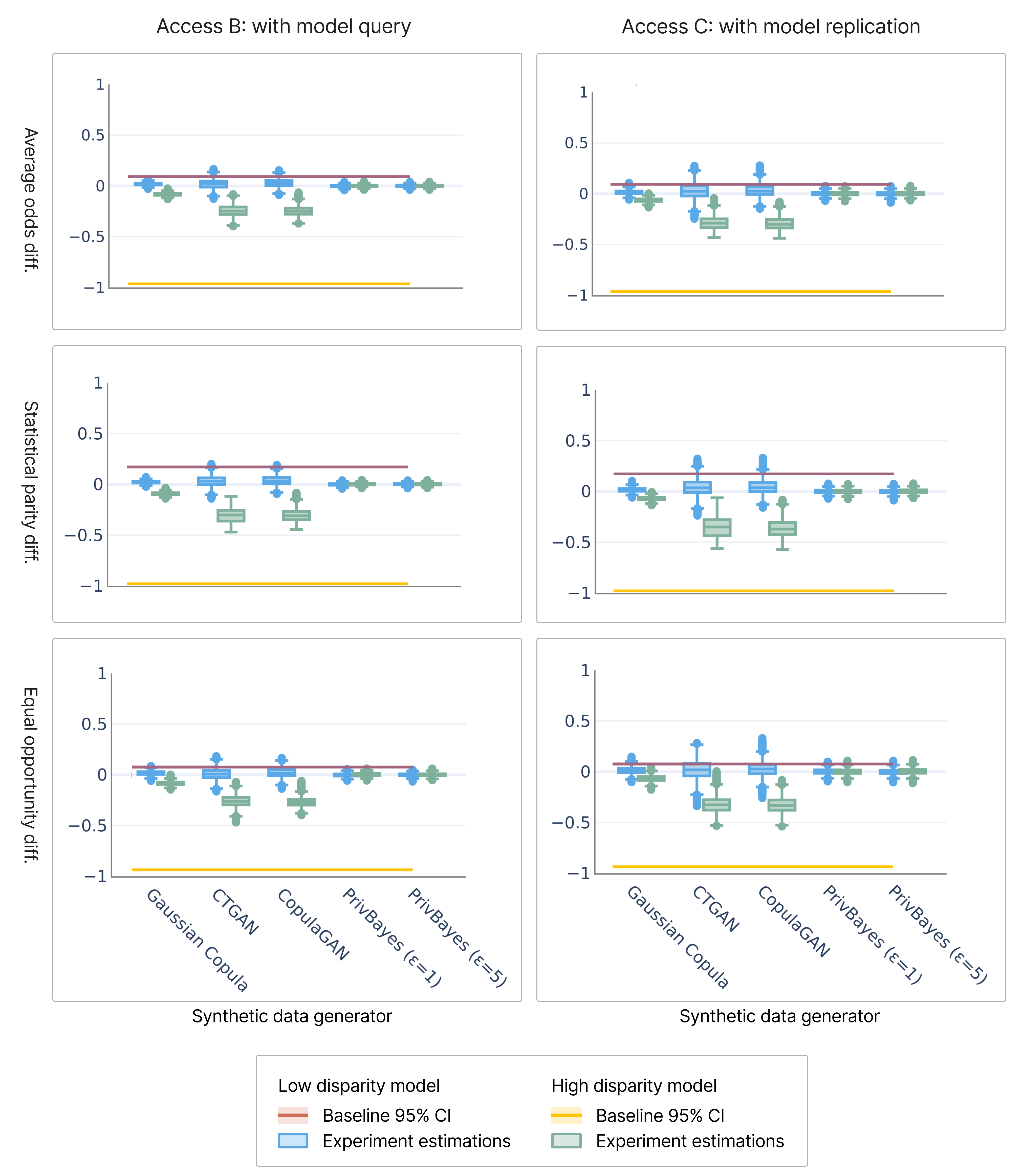}
\caption{\textbf{Metric reliability based on models used for synthetic data generation}, for Scenario B (left) and Scenario C (right) (ACS dataset)}
\label{fig:acs_exp5}
\Description[Box plots for metric estimations based on the model used to generate synthetic data]{The figure includes 2 subplots for each parity metric; one for Access Scenario B (with model predictions) and one for Access Scenario C (with model replication). The synthetic data generation models on the x-axis are ``Gaussian Copula'', ``CTGAN'', ``CopulaGAN'', ``PrivBayes with epsilon=1'', and ``PrivBayes with epsilon=5''. For both low and high disparity cases, none of the audits on synthetic data yield parity values that overlap with the de facto true values.}
\end{figure*}


\end{document}